\documentclass[fleqn,usenatbib]{mnras}

\usepackage{newtxtext,newtxmath}

\usepackage[T1]{fontenc}
\usepackage{ae,aecompl}

\usepackage{graphicx}	
\usepackage{amsmath}	
\usepackage{amssymb}	

\def \apjl {ApJL}

\def \apjs {ApJS}
\def \aap {A\&A}

\def \spose#1{\hbox  to 0pt{#1\hss}}  
\def \lta{\mathrel{\spose{\lower 3pt\hbox{$\sim$}}\raise  2.0pt\hbox{$<$}}}
\def \gta{\mathrel{\spose{\lower  3pt\hbox{$\sim$}}\raise 2.0pt\hbox{$>$}}}

\def \kms {\ifmmode  \,\rm km\,s^{-1} \else $\,\rm km\,s^{-1}  $ \fi }
\def \kpc {\ifmmode  {\rm kpc}  \else ${\rm  kpc}$ \fi  }  
\def \hkpc {\ifmmode  {h^{-1}\rm kpc}  \else ${h^{-1}\rm kpc}$ \fi  }  
\def \hMpc {\ifmmode  {h^{-1}\rm Mpc}  \else ${h^{-1}\rm Mpc}$ \fi  }  
\def \Msun {\ifmmode \rm M_{\odot} \else $\rm M_{\odot}$ \fi}
\def \hMsun {\ifmmode h^{-1}\,\rm M_{\odot} \else $h^{-1}\,\rm M_{\odot}$ \fi}
\def \hhMsun {\ifmmode h^{-2}\,\rm M_{\odot}\else $h^{-2}\,\rm M_{\odot}$ \fi}
\def \Lsun {\ifmmode L_{\odot} \else $L_{\odot}$ \fi} 
\def \hhLsun {\ifmmode h^{-2}\,\rm L_{\odot} \else $h^{-2}\,\rm L_{\odot}$ \fi}

\def\LCDM{$\Lambda$CDM }
\def \LCDM {\ifmmode \Lambda{\rm CDM} \else $\Lambda{\rm CDM}$ \fi}
\def \sig8 {\ifmmode \sigma_8 \else $\sigma_8$ \fi} 
\def \Omegam {\ifmmode \Omega_{\rm m} \else $\Omega_{\rm m}$ \fi} 
\def \Omegab {\ifmmode \Omega_{\rm b} \else $\Omega_{\rm b}$ \fi} 
\def \Omegar {\ifmmode \Omega_{\rm r} \else $\Omega_{\rm r}$ \fi} 
\def \fbar {\ifmmode f_{\rm b} \else $f_{\rm b}$ \fi} 
\def \OmegaL {\ifmmode \Omega_{\rm \Lambda} \else $\Omega_{\rm \Lambda}$\fi} 
\def \Deltavir {\ifmmode \Delta_{\rm vir} \else $\Delta_{\rm vir}$ \fi}
\def \rhocrit {\ifmmode \rho_{\rm crit} \else $\rho_{\rm crit}$ \fi}

\def \rs {\ifmmode r_{\rm s} \else $r_{\rm s}$ \fi} 
\def \rh {\ifmmode r_{\rm h} \else $r_{\rm h}$ \fi} 
\def \Rvir {\ifmmode R_{\rm vir} \else $R_{\rm vir}$ \fi}
\def \Vvir {\ifmmode V_{\rm  vir} \else  $V_{\rm vir}$  \fi} 
\def \Vmax {\ifmmode V_{\rm  max} \else  $V_{\rm max}$  \fi} 
\def \Mvir {\ifmmode M_{\rm  vir} \else $M_{\rm  vir}$ \fi}  
\def \Mhalo {\ifmmode M_{200} \else $M_{200}$ \fi}  
\def \Nvir {\ifmmode N_{\rm  vir} \else $N_{\rm  vir}$ \fi}  
\def \Jvir {\ifmmode J_{\rm vir} \else $J_{\rm vir}$ \fi} 
\def \Evir {\ifmmode E_{\rm vir} \else $E_{\rm vir}$ \fi} 
\def \lam {\ifmmode \lambda  \else $\lambda$ \fi} 
\def \lamp {\ifmmode \lambda^{\prime} \else $\lambda^{\prime}$  \fi} 
\def \lampc {\ifmmode \lambda^{\prime}_{\rm c} \else
  $\lambda^{\prime}_{\rm c}$  \fi} 

\def \xoff {\ifmmode x_{\rm off} \else $x_{\rm off}$ \fi}
\def \rhorms {\ifmmode \rho_{\rm rms} \else $\rho_{\rm rms}$ \fi}
\def \qbar {\ifmmode \bar{q} \else $\bar{q}$ \fi}

\def \Mb {\ifmmode M_{\rm b} \else $M_{\rm b}$ \fi} 
\def \eSF {\ifmmode \epsilon_{\rm SF} \else $\epsilon_{\rm SF}$ \fi} 
\def \Md {\ifmmode M_{\rm d} \else $M_{\rm d}$ \fi} 
\def \Mg {\ifmmode M_{\rm g} \else $M_{\rm g}$ \fi} 
\def \Rb {\ifmmode R_{\rm b} \else $R_{\rm b}$ \fi} 
\def \Rd {\ifmmode R_{\rm d} \else $R_{\rm d}$ \fi} 
\def \Rg {\ifmmode R_{\rm g} \else $R_{\rm g}$ \fi} 
\def \mgal {\ifmmode m_{\rm gal} \else $m_{\rm gal}$ \fi} 
\def \rj {\ifmmode {\cal R}_j \else ${\cal R}_j$ \fi} 
\def \lamgal {\ifmmode \lambda_{\rm gal} \else $\lambda_{\rm gal}$ \fi} 
\def \Vcirc {\ifmmode V_{\rm circ} \else $V_{\rm circ}$ \fi} 
\def \Vrot {\ifmmode V_{\rm rot} \else $V_{\rm rot}$ \fi} 
\def \Vflat {\ifmmode V_{\rm flat} \else $V_{\rm flat}$ \fi} 
\def \Mstar {\ifmmode M_{\rm star} \else $M_{\rm star}$ \fi} 
\def \Mgas {\ifmmode M_{\rm gas} \else $M_{\rm gas}$ \fi} 
\def \Mbar {\ifmmode M_{\rm bar} \else $M_{\rm bar}$ \fi}
\def \Rbar {\ifmmode R_{\rm bar} \else $R_{\rm bar}$ \fi} 

\def \DeltaIMF {\ifmmode \Delta_{\rm IMF} \else $\Delta_{\rm IMF}$ \fi}

\def \VV {\ifmmode V_{\rm 2.2}/V_{200} \else $V_{2.2}/V_{200}$ \fi} 
\def \dvr {\ifmmode \partial_{\rm VR} \else $\partial_{\rm VR}$ \fi}

\title[Origin of bulge stars]{Angular momentum evolution of bulge stars in disc galaxies in NIHAO}

\author[L. Wang et al.]{
Liang Wang$^{1}$\thanks{E-mail: liang.wang@uwa.edu.au},
Danail Obreschkow$^{1}$,
Claudia del P. Lagos$^{1,2}$,
Sarah M. Sweet$^{3}$,
\newauthor{Deanne Fisher$^{3}$,
Karl Glazebrook$^{3}$,
Andrea V. Macci{\`o}$^{4,5}$,
Aaron A. Dutton$^{4}$,
Xi Kang$^{6}$}
\\
$^{1}$ International Centre for Radio Astronomy Research (ICRAR), M468, University of Western Australia, 35 Stirling Hwy, \\ 
Crawley, WA 6009, Australia\\
$^{2}$ ARC Centre of Excellence for All Sky Astrophysics in 3 Dimensions (ASTRO 3D) \\
$^{3}$ Centre for Astrophysics and Supercomputing, Swinburne University of Technology, P.O.Box 218, Hawthorn, VIC 3122, Australia\\
$^{4}$ New York Univsersity Abu Dhabi, P.O.Box 129188, Saadiyat Island, Abu Dhabi, United Arab Emirates\\
$^{5}$ Max-Planck-Institut f\"ur Astronomie, K\"onigstuhl 17, D-69117 Heidelberg, Germany\\
$^{6}$ Purple Mountain Observatory, the Partner Group of MPI f\"ur Astronomie, 2 West Beijing Road, Nanjing 210008, China
}

\date{Accepted XXX. Received YYY; in original form ZZZ}

\pubyear{2018}

\begin{document}
\label{firstpage}
\pagerange{\pageref{firstpage}--\pageref{lastpage}}
\maketitle

\begin{abstract}
We study the origin of bulge stars and their angular momentum (AM) evolution
in 10 spiral galaxies with baryonic masses above $10^{10}\Msun$ in the NIHAO galaxy formation simulations.
The simulated galaxies are in good agreement with observations of 
the relation between specific AM and mass of the
baryonic component and the stellar bulge-to-total ratio ($B/T$).
We divide the star particles at $z=0$ into disc and bulge components using a hybrid photometric/kinematic decomposition method that identifies all central mass above an exponential disc profile as the ``bulge''. By tracking the bulge star particles back in time, we find that on average 95\% of the bulge stars formed {\it in situ}, 3\% formed {\it ex situ} in satellites of the same halo, and only 2\% formed {\it ex situ} in external galaxies. The evolution of the AM distribution of the bulge stars paints an interesting picture: the higher the final $B/T$ ratio, the more the specific AM remains preserved during the bulge formation. In all cases, bulge stars migrate significantly towards the central region, reducing their average galactocentric radius by roughly a factor 2, independently of the final $B/T$ value. However, in the higher $B/T$ ($\gtrsim0.2$) objects, the velocity of the bulge stars increases and the AM of the bulge is almost conserved, whereas at lower $B/T$ values, the velocity of the bulge stars decreases and the AM of bulge reduces.
The correlation between the evolution of the AM and $B/T$ suggests that bulge and disc formation are closely linked and cannot be treated as independent processes.
\end{abstract}




\section{Introduction}

Angular momentum (AM) $J$ has long been recognized as a fundamental quantity in galaxy formation and evolution.
It is often convenient and physically meaningful to remove the implicit  mass scaling of $J$ by adopting the specific AM $j \equiv J/M$.
The standard picture of cosmic structure formation predicts that dark matter (DM) and baryonic gas acquire similar specific AM via tidal torques from neighbouring structures. In the simple model, where the gas cools radiatively and settles in a spinning disc, the size of this disc is proportional to its $j$  \citep{fall80}, whereas the size of the DM halo is independent of AM.


Empirically, the global morphology of galaxies, as quantified by Hubble's classical tuning fork, correlates strongly with specific AM. In fact, early-type and 
late-type galaxies at redshift $z = 0$ obey similar scaling relations between stellar specific AM $j_{\star}$ and mass $M_{\star}$, of the form 
$j_{\star} \propto M^{\alpha}_{\star}$ with $\alpha \sim 2/3$ \citep{fall83,romanowsky12,fall13,obreschkow14,cortese16}. However, these relationships are offset by about a factor $\sim 5$ in $j_\star$, with early-types having less AM. \citet{obreschkow14} refine this picture by using high-precision AM data from the THINGS survey and optical $K$-band bulge-to-total ratios ($B/T$). In a sample with $0<$ $B/T$ $\lesssim 0.3$ and baryon mass $M_b$ (stellar and cold gas mass) in the range $10^9 \Msun < M_b < 10^{11} \Msun$, they find a very strong correlation between $M_b$, the baryon specific AM $j_b$ and $B/T$.

It is a pressing theoretical challenge to explain how this observed relationship between morphology and AM arises in regular galaxies. This work aims to shine light on this discussion by tackling the time-evolution of the bulge-stars in regular star-forming disc galaxies in a cosmological smoothed-particle hydrodynamic (SPH) simulation. Naturally, this goal requires a clear definition of the `bulge', not just in terms of its global mass and extent, but in terms of identifying all star particles of the bulge, so that they can be traced back in time. Observations regularly use photometric bulge-disc decomposition methods, relying on surface density (defining the bulge as a central overdensity) and/or colour (defining the bulge as the old central population). In turn, simulations often employ kinematic approaches (defining the bulge as the low-angular momentum component, e.g.\ \citet{obreja16}). These definitions result in different bulge mass fractions, and, more importantly, in a very different sets of stars labelled as bulge-stars. To do justice to this complexity, we first carry out a detailed comparison of photometric and kinematic bulge-disc decompositions (Section 2) and then pick the definition most adequate to the \textit{particular} set of questions addressed in this work.

Cosmological hydrodynamic simulations of galaxies from disc-dominated to bulge-dominated objects compare the the simulation results with observations by considering the Tully-Fisher relation \citep{robertson04,governato07,stinson10,piontek11,dutton17}, the AM content \citep{scannapieco09} and disc sizes \citep{brook11}.
However, the simulated galaxies from early studies always had unrealistic massive and concentrated bulge to the corresponding observed bulge \citep{binney01,bullock01,bosch01,bosch02,donghia06,donghia07,stinson10,scannapieco12}. The failure of reproducing realistic disc galaxies with small bulges comes from the fact that these simulations were plagued by over-cooling and AM losses resulting in compact, bulge-dominated galaxies.

In the past years, cosmological simulations have reached enough resolution to resolve bulges and included more realistic models for star formation and stellar 
feedback. These improvements made it possible to produce realistic galactic discs \citep{guedes11,agertz11,aumer13,stinson13b,marinacci14,roskar14,murante15,colin16,grand17}.
The most recent generation of large-volume hydrodynamical simulations of galaxy formation, e.g. \citet{vogelsberger14,schaye15} managed to reproduce
observed galaxies from disc-dominated sequence and bulge-dominated sequence \citep{genel15,teklu15,zavala16,lagos17,sokolowaska17,defelippis17}.

In the conventional view, bulge stars have two main origins: galaxy-galaxy mergers and {\it in situ} formation in galactic discs. The former tend to result in pressure supported, ``classical'' bulges \citep{davies83}, while the latter give rise to rotation-supported ``pseudo''-bulges\citep{kormendy04}. \citet{donghia07} show that cooling in the highly dense inner region and dynamical friction of satellite-galaxies dissipate the AM of gas resulting in compact discs.
In contrast, \citet{springel05} and \citet{robertson06} show that in idealized merger simulations with strong stellar feedback it is possible to form a disc-dominated galaxy,
if the initial disc is gas rich. Recent cosmological simulations also confirm that strong stellar feedback is able to reduce
the bulge mass and result in disc-dominated galaxies \citep{governato07,guedes11,brook12,martig12,stinson13,kannan14,marinacci14,christensen14}.
Furthermore, \citet{hopkins09} and \citet{kannan15} suggest that mergers are not as efficient as previously thought in transforming discs into bulges and bulge formation depends on the gas disc fraction. \citet{lagos17b} using the EAGLE simulations divided mergers into dry (gas-poor) and wet (gas-rich), major/minor and analysed different spin alignments and orbital parameters. They found the mergers' influence on galaxy spin can be quite varied, and in the case of gas-rich mergers, galaxies can significantly spin up, while dry
mergers tend to spin galaxies down. Therefore, the bulge formation couples with complex and only partially understood small-scale processes.

In this paper, we examine the origin and AM evolution of bulge stars in a subset of 10 spiral galaxies from the Numerical Investigation of a Hundred 
Astrophysical Objects, NIHAO \citep{wang15} project. All these bulges have pseudo-bulge properties. The NIHAO simulations are a suite of 88 hydrodynamical cosmological zoom-in
simulations implementing the tree-smoothed particle hydrodynamics (SPH), GASOLINE.
The simulated galaxies cover a range in mass from $10^5 \lesssim M_\star / \Msun \lesssim 10^{11}$.
The prescription for star formation and stellar feedback was developed in the ``Making Galaxies in a 
Cosmological Context'' (MaGICC, \citealp{stinson13}) project. The NIHAO runs keep the
same stellar physics at all scales. The stellar mass of each halo in the NIHAO sample agrees with the prediction
from abundance matching \citep{wang15}.
The galaxies in the NIHAO sample reproduce several observational results, such as
the star formation main sequence \citep{wang15}, the column density profile of cool 
HI \citep{gutcke17} and the Tully-Fisher relation \citep{dutton17}. 
Given its high mass resolution, 
this sample allows us to track the evolution of AM of bulge star particles
and witness the formation of a bulge in a realistic cosmological environment.

This paper is organized as follows: In Section \ref{sec:sim}, we give an overview of the
cosmological hydrodynamical simulations and the post-processing applied in this work. 
In Section \ref{sec:formation}, we study the origin of bulge stars and discuss the
evolution of their AM, locations and velocities. Section \ref{sec:discuss} discusses the coherent picture of how the bulges of 
galaxies form. Finally, in Section \ref{sec:summary} we discuss and summarize our conclusions.

\section{Simulations} \label{sec:sim}

The simulations we analyse in this work are a subsample of the NIHAO suite 
\citep{wang15}, based on updated version of the MaGICC project 
\citep{stinson13}. 
The NIHAO simulation suite is an unbiased sample of 
isolated haloes of 
present-day masses between $M_{\rm halo} \sim 4 \times 10^9 \Msun$ and 
$M_{\rm halo} \sim 4 \times 10^{12} \Msun$. 
The initial conditions are created so that 
we typically have a million DM particles 
inside the virial radius of the target halo at redshift $z = 0$, across the whole mass range.
We adopted the latest compilation of cosmological parameters from the
Planck satellite \citep{planck14}.
DM particle masses range from $\sim 10^{4} \Msun$ in  our lowest
mass haloes to $\sim 10^{6} \Msun$ in our most massive  haloes, and
their force softenings range from $\sim$ 150 pc to $\sim$ 900 pc
respectively. Gas particles are less massive by factor of
$(\Omega_{\rm DM}/\Omega_{\rm b})\simeq 5.48$, and  the
corresponding force softenings are 2.34 times smaller. 
A comparison with other state-of-the-art simulations can be found
in Fig. 2 in \citet{wang15}. More
information on the collisionless parent simulations, the force
softenings and particle masses for the highest refinement level for
each simulation and sample  selection can be found in \citet{Dutton14}
and \citet{wang15}.

We use the SPH hydrodynamics code {\sc gasoline} \citep{Wadsley04},
with a revised treatment of  hydrodynamics as described in
\citet{Keller14}. The code includes a subgrid model for turbulent
mixing of metal and energy \citep{Wadsley08}, heating and cooling
include photoelectric heating of dust grains, ultraviolet (UV) heating
and ionization and  cooling due to hydrogen, helium and metals
\citep{Shen10}.

\subsection{Star formation and feedback}
The star formation and feedback modelling follow what was used in the 
MaGICC simulations \citep{stinson13}.
The star formation recipe adopts a Kennicutt-Schmidt relation, where the 
star-forming gas has a temperature lower than 15000 K and a density higher
than 10.3 ${\rm cm}^{-3}$. 
Supernova feedback is implemented following the blast-wave formalism
\citep{stinson06}. 
Another stellar feedback mechanism ejects energy prior to supernovae
explosions which accounts for the photoionizing radiation of massive
stars \citep{stinson13}.
The free parameters in the
feedback scheme were chosen such that a MW mass galaxy fits the stellar
mass-halo mass relation at $z = 0$. The NIHAO simulations form the right
amount of stars and cold gas as evidenced by the consistency with the stellar
mass versus halo mass relations from abundance matching since $z = 4$
\citep{wang15} and the cold gas versus stellar mass relation at $z = 0$
\citep{stinson15}.

\subsection{Bulge-disc decomposition} \label{sec:decomposition}

\begin{figure*}
    \includegraphics[width=0.40\textwidth, trim={0.6cm 0 1.2cm 0},clip]{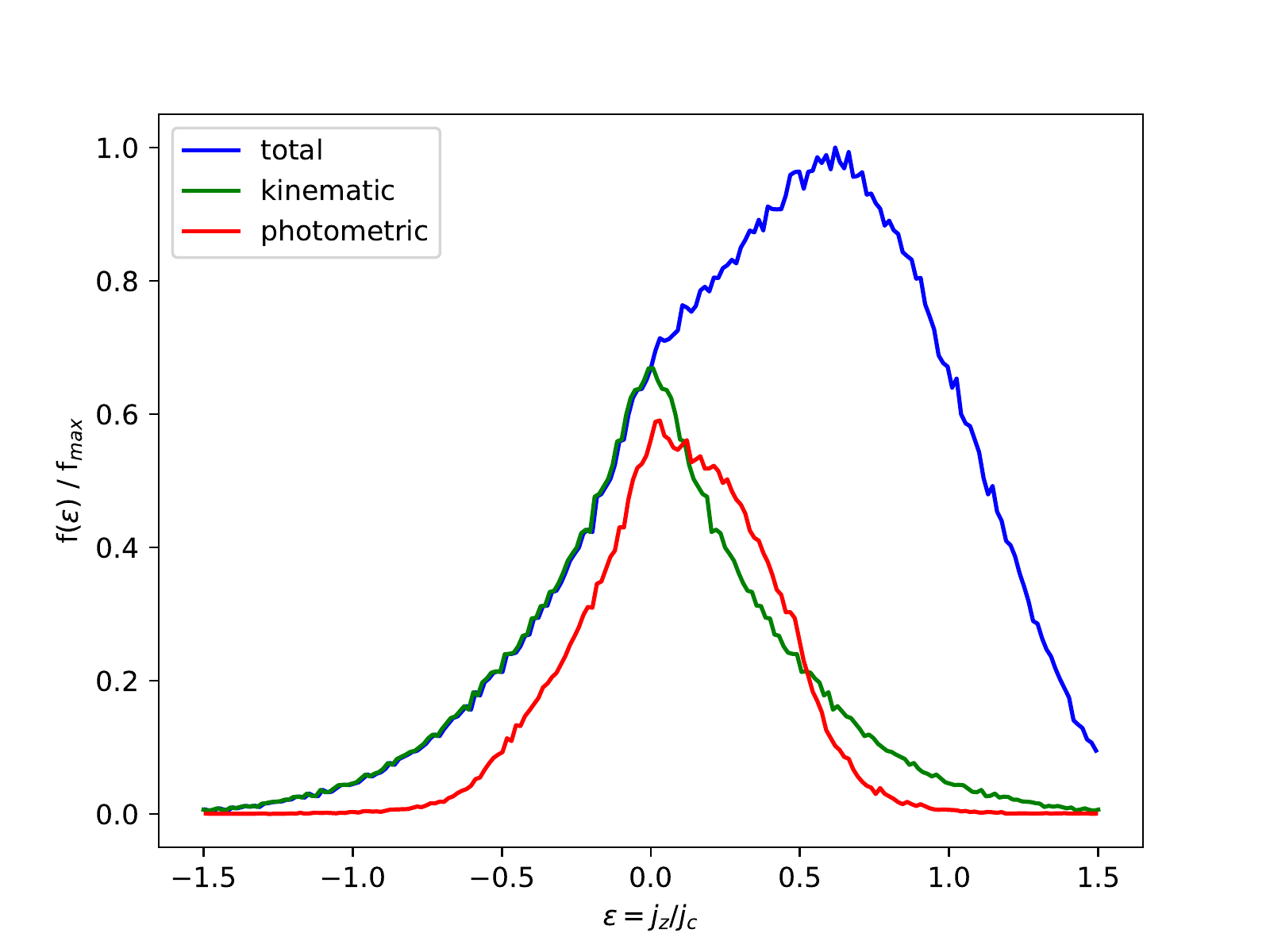}
    \includegraphics[width=0.59\textwidth, trim={0.6cm 0 1.8cm 0},clip]{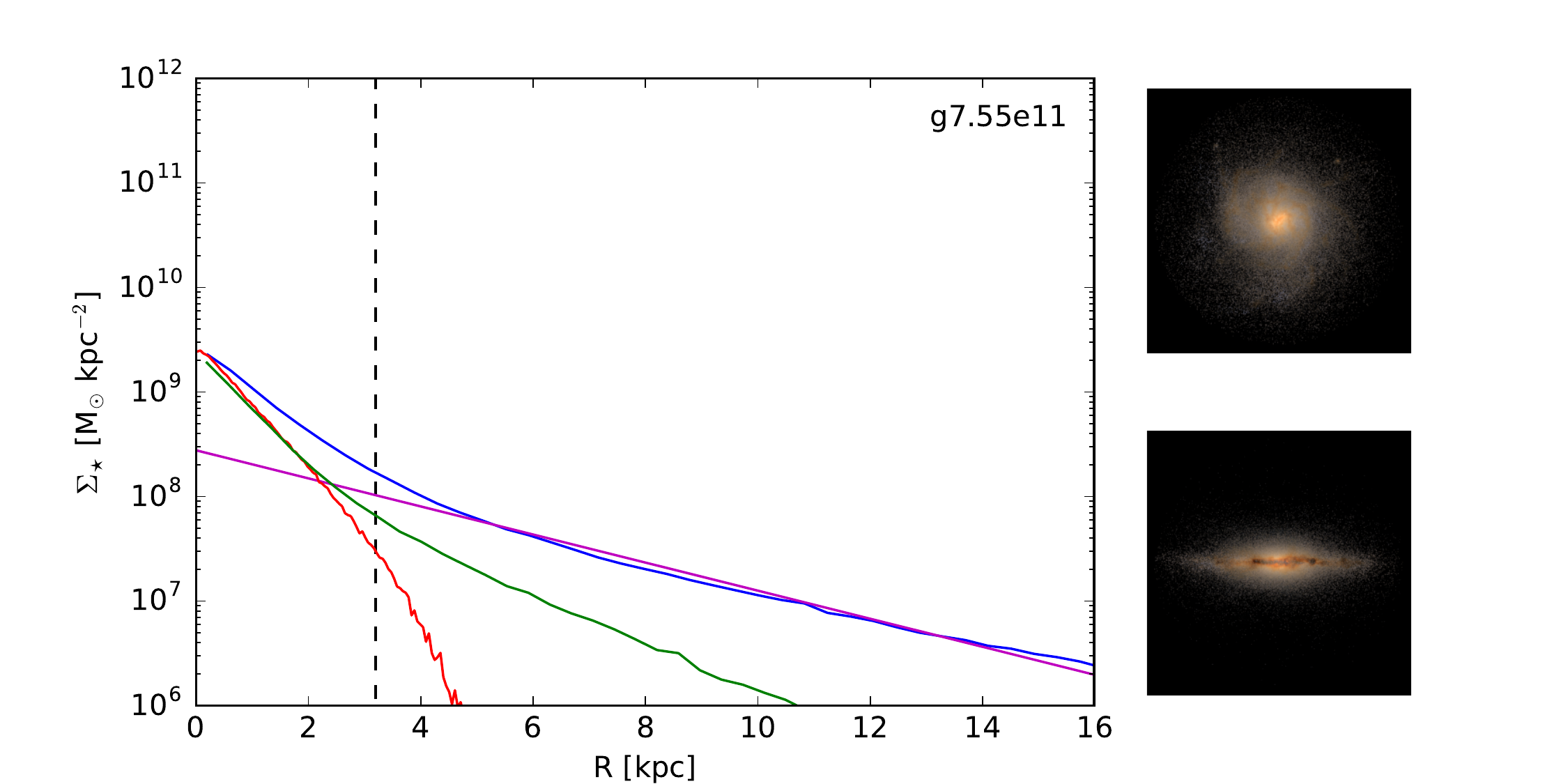}
    \caption{Left: The distribution of the circularity parameter $\epsilon$ for all stellar particles within five half mass
         radii of galaxy g7.55e11 (blue line), bulge particles selected by kinematic decomposition (green line) and bulge particles
         selected by photometric decomposition (red line).
         Right: Stellar surface density profiles of same galaxy (blue filled circle) at z = 0. The magenta line shows the exponential fit
         of the disc component and the dashed line shows the half mass radius. The images illustrate the face-on 
         and edge-on synthetic images of the galaxies after processing them via the Monte Carlo radiative transfer code 
         {\sc SUNRISE}\citep{jonsson06}.
         All images measure 50 kpc on a side.
         \label{fig:pro_img}}
\end{figure*}

\begin{figure}
	\includegraphics[width=\columnwidth]{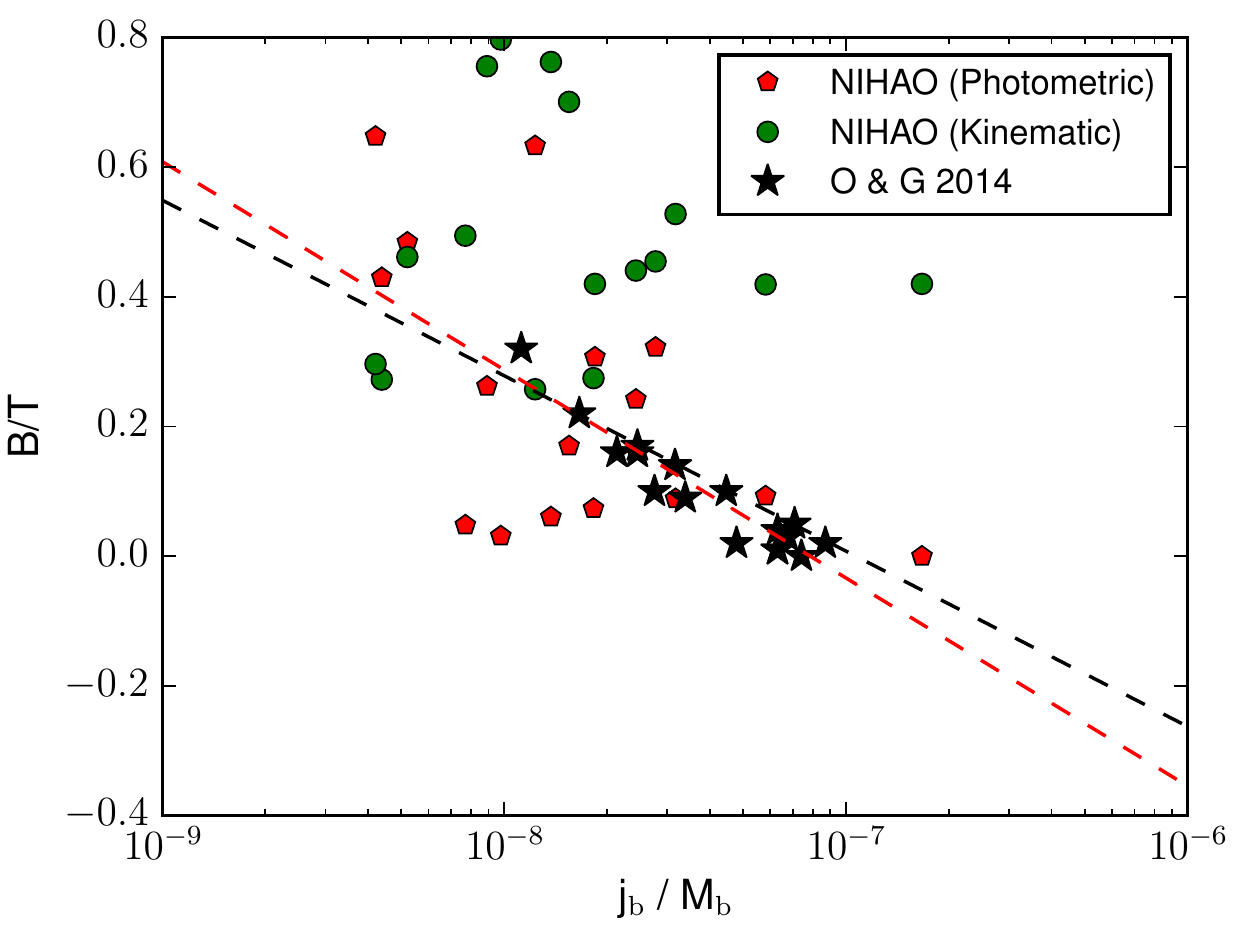}
    \caption{Representation of the simulated galaxies and THINGS observations \citep{obreschkow14} in the ($j_b/M_b,B/T$)-plane.
         The red and black dashed lines are linear fits to both datasets.
         \label{fig:jbmb_beta}}
\end{figure}

We decompose our simulated galaxies at $z=0$ into disc and bulge components. 
We perform these decompositions considering only
the star particles inside the spherical region within 20\% of the
virial radius, 
with the centre of the halo being placed at the centre of mass of the target galaxy. We orient 
the axes such that the $z$-axis is parallel to the direction of the angular momentum vector of the galaxies.
We do not identify and remove particles that might belong to a stellar 
halo as most of the stars within this radius actually belong to the disc-bulge
system and decomposition results are not sensitive to the halo stars. 

Many methods have been presented to perform bulge-disc decompositions in observed and simulated data sets. Importantly, these methods correspond to different definitions of a ``bulge'' and hence it is important to pick the definition and then the method most adequate for the questions to be answered. To illustrate the importance of choosing a method, we first compare two standard but very distinct methods: a kinematic method based on the particle orbits and a photometric method based on the surface density.

Kinematic decompositions of stellar system have evolved over the last decade. \citet{abadi03} introduced a one-dimensional decomposition based on the circularity parameter of stellar orbits. The circularity parameter $\epsilon$ is defined as $j_z/j_c$,
where $j_z$ is the $z$-component of the AM vector of a stellar particle when the galactic disc lies in the x-y plane, and $j_c$ is the norm of the specific angular
momentum vector of a hypothetical particle at the same location on a circular orbit.
In this work, we define $j_c = \sqrt[]{GMr}$ where $r$ is the radius of the particle.
The higher the circularity, the more the particle's motion corresponds to a disc-like orbit. 
The kinematic method does a good job determining the relative mass of disc and spheroid (see also \citet{scannapieco11, martig12}). It assumes that the spheroid is non-rotating and thus has a symmetric distribution of circularities around $\epsilon \sim 0$. The thin disc gives the sharp peak at circularities close to $\epsilon \sim 1$. We identify the spheroidal component by this assumption, so that its mass is $M_{\rm spheroid} = 2 \times M(\epsilon < 0)$.

Secondly, we decompose galaxies into discs and bulges photometrically following the observational approach of \citet{obreschkow14}.
We generate a face-on image of the stellar component and extract one-dimensional stellar surface density profiles for each galaxy.
Explicitly, we bin the particles by their cylindrical radii and divide the mass in each bin by the surface area of the annulus.
The stellar surface density profiles hence obtained are then fitted by an exponential function
\begin{equation}
	\Sigma_{\rm d} \left( r \right) = \Sigma_{\rm d,0} \exp \left( -\frac{r}{R_{\rm d}}
\right) 
\end{equation}
where $R_{\rm d}$ is the scale radius of the profile and $\Sigma_{\rm d,0}$ is 
the central surface density. We determine these two parameters with a least-squares
minimization, excluding the inner half-mass radius to avoid fitting the exponential to the bulge.


Using the disc fits, we then identify the bulge stars.
To do so, we bin the galaxy into 100 radial bins, each corresponding to an annulus in the face-on projection. In each annulus $i$, the mass fraction $f_i$ of the stars in the bulge is calculated by
\begin{equation}
f_i = \frac{\Sigma_{{\rm total},i}-\Sigma_{{\rm d},i}}{\Sigma_{{\rm total},i}},
\end{equation}
where $\Sigma_{\rm total,i}$ is the mean stellar surface density in the annulus and $\Sigma_{\rm d,i}$ is the disc surface density of the fit $\Sigma_{\rm d} \left( r \right)$, evaluated at the central radius of the annulus. This definition of $f_i$ is only adopted out to the radius, where the total surface density first falls below $\Sigma_{\rm d} \left( r \right)$. Beyond this radius, we set $f_i=0$ to avoid including spiral
arms and halo stars in the bulge mass. The global $B/T$ ratio is then defined as $B/T=M_{\star}^{-1}\sum_i{f_i m_i}$, where $m_i$ is the total stellar mass in the annulus $i$ and $M_{\star}$ is the total stellar mass in the galaxy.

We use kinematic criterion to decide which of the star particles in each annulus are bulge star particles.
We therefore label, in each annulus $i$, the particles with the
lowest circularities as bulge star particles, until they make up a mass fraction $f_i$.
An illustration of the bulge particles identified in this way is shown in
Fig.~\ref{fig:gasmap} (coloured points and yellow stars), discussed later.

Fig.~\ref{fig:pro_img} shows both the kinematic and photometric decompositions of the galaxy g7.55e11. The kinematic decomposition relies on the circularity distribution shown in the left panel: by definition, the negative part of this distribution and a mirrored positive part (green line) is said to be the ``kinematic bulge''. In turn, the ``photometric bulge'' is identified as the excess above an exponential surface density profile (red line in right panel). The circularity distribution and surface densities of all stars are shown as blue lines in both panels, while the kinematic and photometric bulges are shown as green and red lines, respectively. The right-most small panels show the face-on and edge-on views of the galaxy after processing it through the radiative transfer code {\sc SUNRISE} \citep{jonsson06}. In this discy galaxy the circularity distributions of the kinematic and photometric bulges are similar. However, their surface density profiles differ significantly. In fact, the kinematically identified bulge extends to almost three half-mass radii, which is difficult reconcile with the standard conception of bulges, but instructive in that it shows that pressure support is relevant even at such large scales. Most other galaxies show more disparity between the kinematic and photometric decompositions (see Fig.~\ref{fig:pro_img_append}).

Fig.~\ref{fig:jbmb_beta} shows a comparison between our simulated galaxies which have bugles decomposed kinematically (green) and photometrically (red) from our sample to those in \citet{obreschkow14} adopting a natural 2-D projection of the $M_b$-$j_b$-$B/T$ correlation that approximately minimises the scatter.
The baryons in the galaxies are defined as the total mass of the stars and the cold gas 
($T<10^4$ K) within 0.2$\Rvir$.
The specific AM vector of these baryons is calculated as
\begin{equation}
	j_b = \frac{|\Sigma m_i \mathbf{r}_i \times \mathbf{v}_i |}{\Sigma m_i},
\end{equation}
where $\mathbf{r}_i$ are the distance vectors from the centre of mass and $\mathbf{v}_i$ are the velocities in the the center of mass frame.
The linear fits shown as red (corresponding to photometric decomposition) and black lines (corresponding to THINGS observation) show a good agreement. However, the NIHAO galaxies with photometric decomposition yield a larger scatter than the observations, which may be attributed to the observational selection of regular gas-rich field galaxies.
The NIHAO galaxies with kinematic decomposition are clearly offset, and show much less correlation between $B/T$ and $j_b / M_b$. The Pearson correlation coefficient of galaxies with photometric decomposition is -0.50, while that of galaxies with kinematic decomposition amounts to an insignificant +0.06.

This discrepancy mainly comes from the difference in the decomposition methods. As described above, the kinematic bulges have a symmetric circularity distribution and are hence non-rotating, on average. However, the empirical relation between $B/T$ and $j_b / M_b$ (stars in Fig.~\ref{fig:jbmb_beta} ) relies on photometrically identified bulges, which turn out to be mostly ``pseudo-bulges'' (\citet{obreschkow14}) that can have some rotational support. Therefore, it is no surprise that the photometric $B/T$ matches the THINGS observations better than the kinematic $B/T$ ratio.

A priori, there is no objective definition of bulges, and different definitions are used for different purposes in the literature \citep{kormendy77, kautsch06, lackner12, fabricius14}. The purpose of this work is to explore the ``bulges'' defined as central overdensities, i.e.~the photometric $B/T$ values. It is likely that this photometric definition of bulges picks out a large number of pseudo-bulges, which make up the majority of the bulges in spiral galaxies in the local Universe \citep{fisher11}. Although such pseudo-bulges are often described as subsystems of galactic discs, it is by no means trivial to understand how these bulges evolve and why their mass correlates with global disc properties -- the topics that we will focus on in the remainder of this paper. Henceforth, ``bulge'' will refer to the photometrically identified bulge, whether it has classical or pseudo-bulge-like properties.

\subsection{$j-M$ distribution} \label{sec:mj_relation}

\begin{figure}
	\includegraphics[width=\columnwidth]{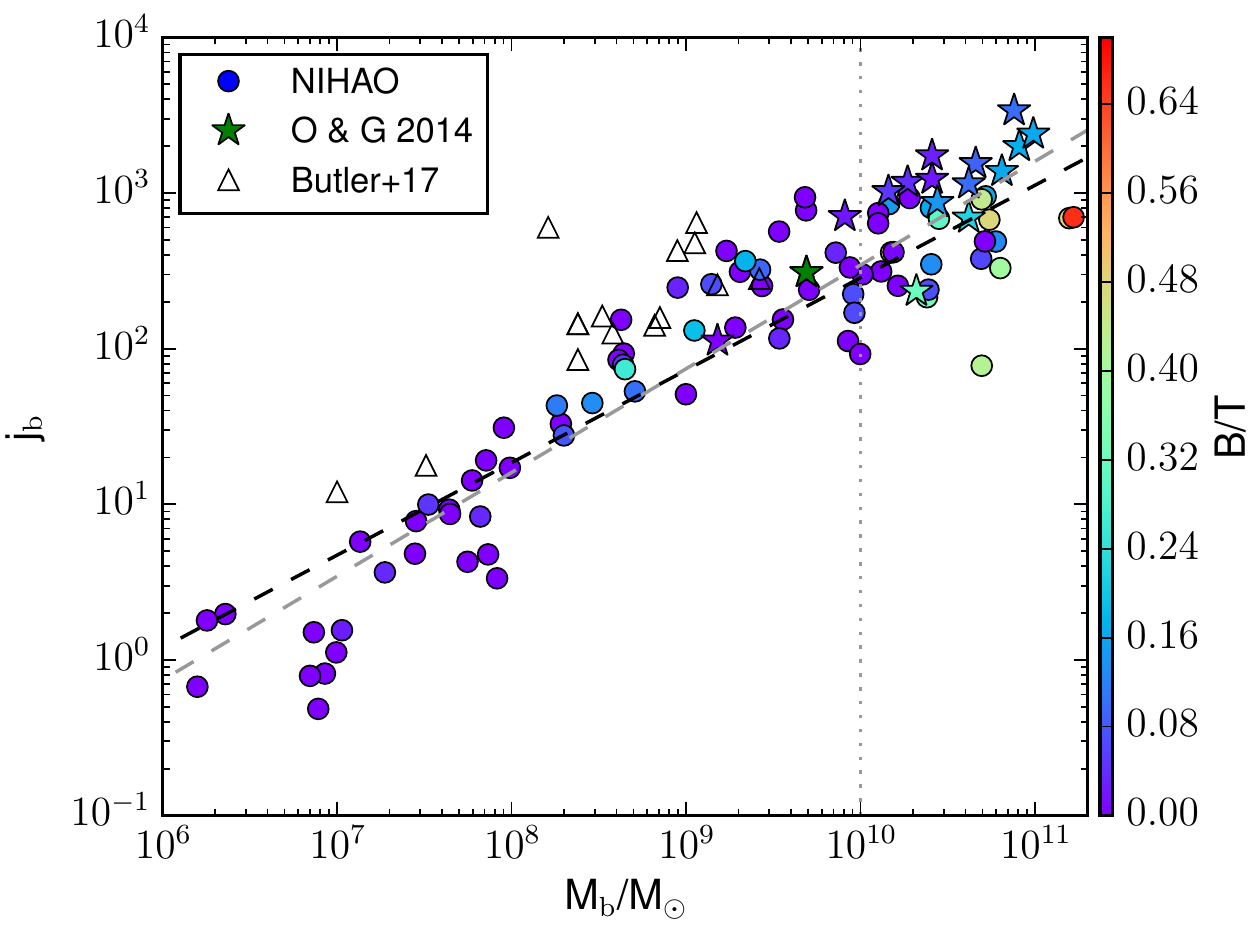}
    \caption{Specific AM--mass diagram ($j_b-M_b$)
         for baryons of simulated galaxies, THINGS observations
         \citep{obreschkow14} and LITTLE THINGS observations \citep{butler17}. 
         The black dashed line is the best fit for simulated galaxies and the grey dashed line is the fitting with a
         fixed slope 2/3.
         \label{fig:jb_mb}}
\end{figure}

\begin{figure}
	\includegraphics[width=\columnwidth]{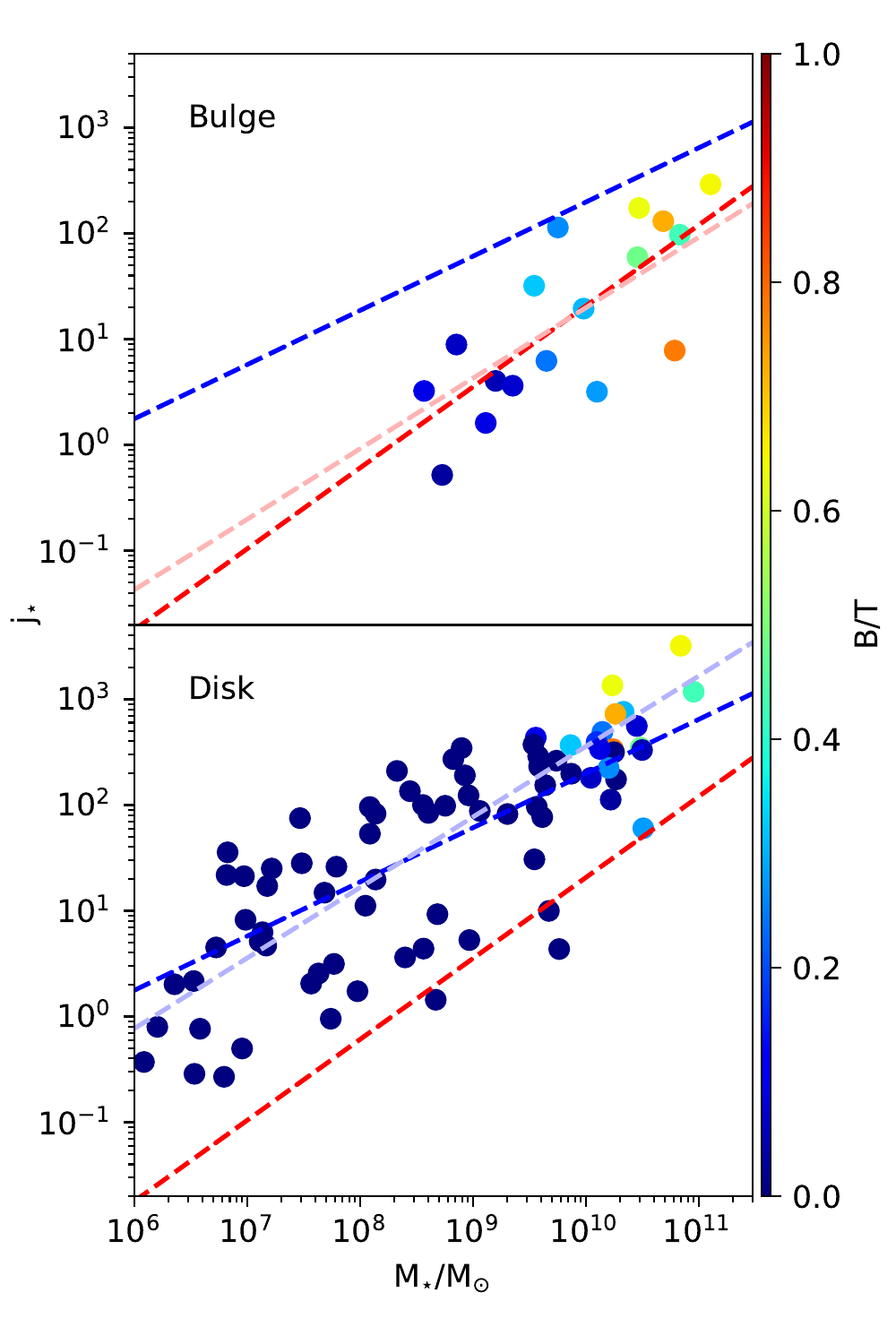}
    \caption{Representation of the simulated galaxies in the mass-AM plane, split into disc 
	(lower panel) and bulge (upper panel) stars. The blue and red dashed lines are the best fits for the disc and bulge components, when using a free slope. The light blue and light red dashed lines are corresponding fits with a fixed 2/3 slope. The points are color-coded by $B/T$.
         \label{fig:b_d}}
\end{figure}

In this section, we analyse the NIHAO galaxies to determine the 
specific AM $j_b$ contained in their baryons at $z=0$, in order to compare the scaling relation between $M_b$ and $j_b$ to observations.

In Fig.~\ref{fig:jb_mb}, our results are plotted against observed galaxies
from the THINGS and LITTLE THINGS analyses \citep{obreschkow14,butler17}.
The location of the simulated galaxies on the diagram is similar to the observed sample.
The points in Fig.~\ref{fig:jb_mb} are colour-coded according to the 
$B/T$, which we deliberately calculated in an analogous way to the observations, as described in section~\ref{sec:decomposition}.
We perform the fits with the HYPER-FIT package \citep{robotham15}.
The best fit to the simulated galaxies has a slope ($\alpha=0.59 \pm 0.03$) similar to the
simplistic analytical prediction of $\alpha = 2/3$ \citep{fall83}.
The difference between simulations and observations might be accounted for by the observational bias of selecting only HI rich objects (especially for LITTLE THINGS) with small bulge components. We refer the reader to \cite{wang18}, where we present a detailed analysis of the connection between gas fraction and angular momentum in NIHAO galaxies. 
The second reason can be the simulations also include a small fraction of ionized gas in the cold gas, they might slightly
overestimate of the observable baryon mass \citep{stinson15,sales17}. The last but not the least, \citet{elbadry18} studied a sample of zoom-in galaxies and compared the AM measured with rotation curve of HI and simply vector-summing AM. The comparison shows, for low mass galaxies, the approach that measures AM by HI rotation curve overestimate galaxies' stellar AM as those stars are primarily supported by dispersion. The discrepancy also consists with this argument.

\citet{obreschkow14} found that the 16 regular spiral galaxies from the THINGS survey display a tight correlation between $M_b, j_b$ and $ B/T$.
The lower specific AM of the simulated galaxies with higher
bulge-to-total ratio is consistent with the empirical finding that the disc and bulge components lie roughly on parallel sequences in the $j$--$M$ plane, separated
by a factor of $\sim 5$ in $j$ \citep{fall13}. Fig.~\ref{fig:b_d} 
compares the individual stellar components, bulge (upper panel) 
and disc (lower panel), in the $j_{\star}$--$M_{\star}$ diagram.
The best fitting lines for each component again show a 
slope similar to $\alpha=2/3$, and the ratio of specific AM between the discs and the bulges range from 4 to 
10 across the 10 galaxies.
The scatter in $j_{\star}$ is about 0.6\ dex (standard deviation) for both components.
The baryonic $j$--$M$ diagram confirms that the simulated galaxies
do not suffer from the overcooling problem or the AM catastrophe
\citep{navarro94,navarro01}, and can be regarded as good laboratories for in-depth 
studies of the AM evolution.

The main purpose of this work is to understand the origin of bulge
stars in disc-dominated galaxies. To do so we explicitly exclude bulge-dominated
objects, i.e.\ we only retain galaxies with $B/T<0.5$. Galaxies with baryon
masses below $10^{10}\Msun$ often admit no bulges or these bulges are very
small and hard to detect, also in the simulation. Such low-mass galaxies also
tend to be cold gas dominated systems in which stellar mass scalings become
more scattered. We therefore chose to limit the analysis to galaxies with
$M_b>10^{10} \Msun$. The resulting sample contains 10 galaxies.

\subsection{Bulge properties}

\begin{figure}
	\includegraphics[width=\columnwidth]{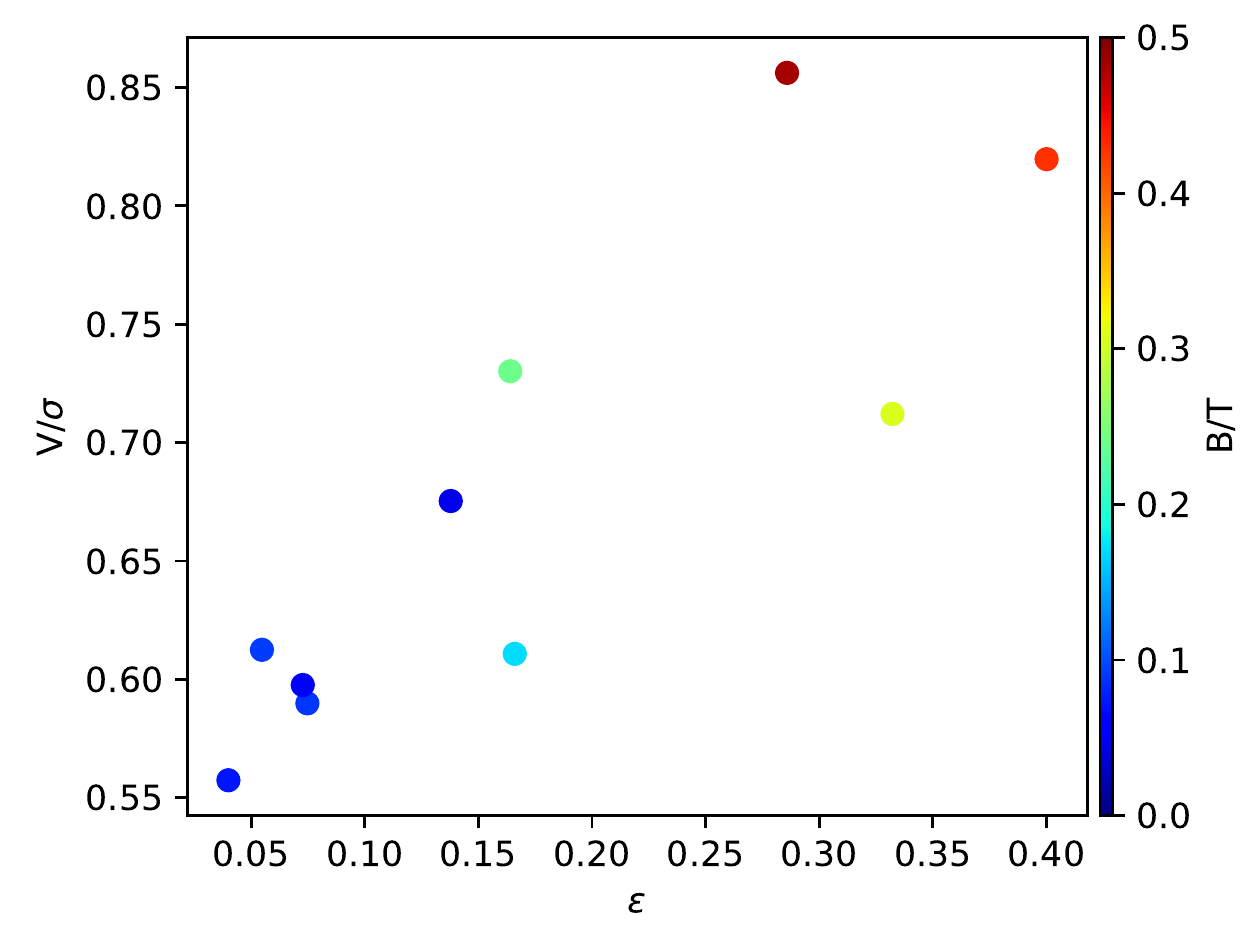}
    \includegraphics[width=\columnwidth]{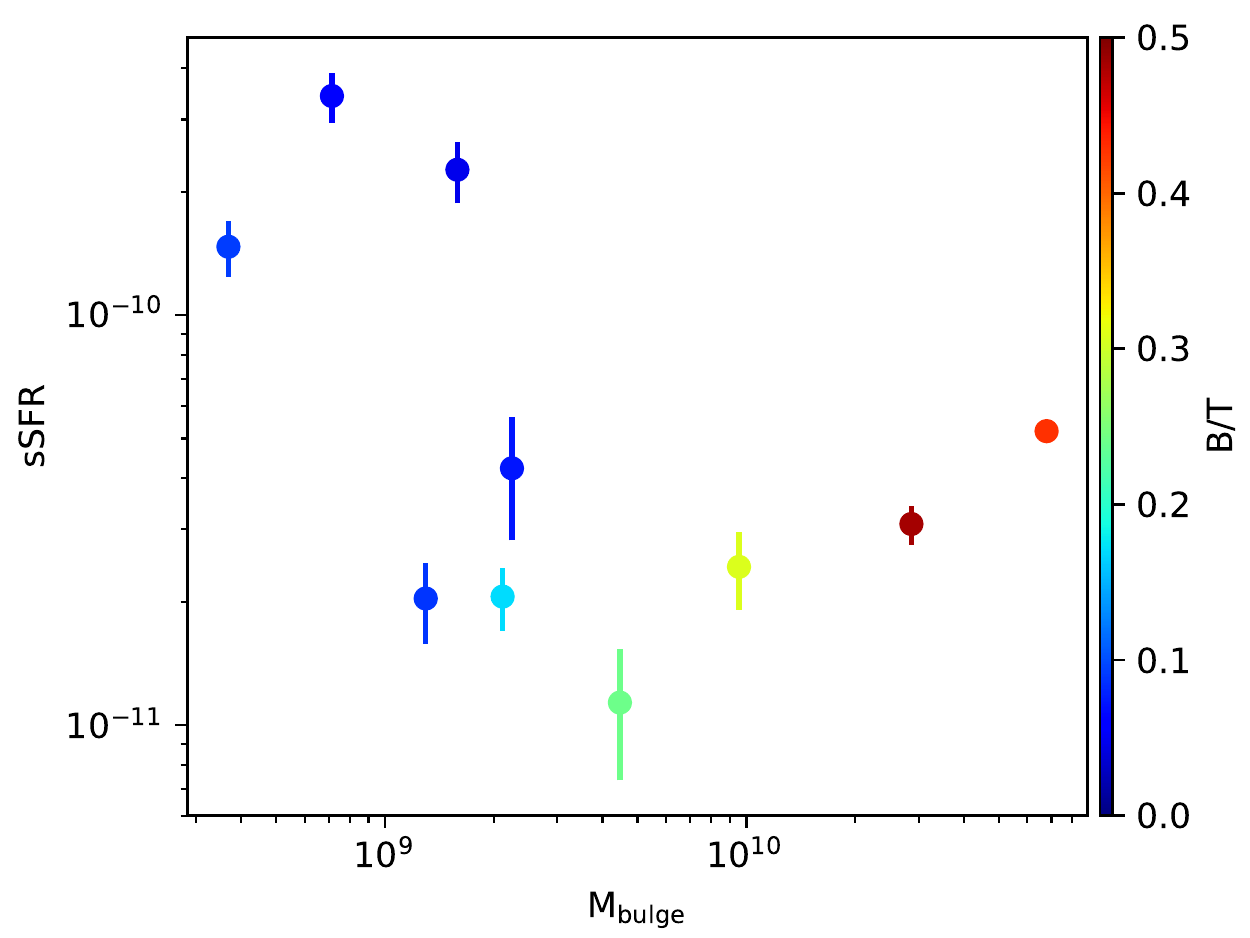}
    \caption{The relative dynamic importance of rotation ($V/\sigma$) as function of 
	 ellipticity for all bulge stars of each galaxy at $z=0$ (upper panel)
         and specific star formation rates as a function of bulge mass (lower panel).
         The points are colour-coded by $B/T$ ratio.
         The error bars in the lower panel represent the shot noise of 
         specific star formation rates for each galaxy.}
         \label{fig:pseudo}
\end{figure}

In Fig.~\ref{fig:pseudo}, the upper panel shows $V/\sigma$ versus ellipticity
$\epsilon$ of the bulge stars at $z=0$ for each galaxy. The rotational velocity $V$ is the
mass-weighted mean of the tangential velocities (i.e.\ in the plane of the disc). 
The velocity dispersion $\sigma$ is the mass weighted mean of $(v_R^2+v_z^2)$, where $v_R$ is the cylindrical radial velocity (in the plane) and $v_z$ is the vertical (parallel to the rotation axis). The ellipticities are calculated as $\epsilon = 1 - \lambda_{min}/\lambda_{mean}$ where $\lambda_{min}$ is the minimum 
eigenvalue of bulge's moment of inertia tensor and $\lambda_{mean}$ is the mean of the other two eigenvalues.
Fig.~\ref{fig:pseudo} shows a positive correlation between $V/\sigma$ and $\epsilon$, meaning that more rotation supported bulges tend to be flatter.
These quantities also correlate with $B/T$, in the sense that larger bulge mass factions correspond to flatter and more rotation-supported bulges.
The maximum ellipticity in the sample is around 0.4, which is still quite spherical (axes ratio about 0.6). This galaxy (g1.92e12) has an extended central overdensity which contains amounts of particles with $\epsilon \sim 1$ so that the $V/\sigma$ of this bulge can be more than $0.80$.

The lower panel in Fig.~\ref{fig:pseudo} shows the specific star
formation rate of the bulge as a function of bulge mass. 
In the simulations, star formation rates are calculated as the total stellar mass formed in the last 100 Myrs, divided by that 
time interval.
On average, the specific star formation rates of bulges in disc dominated
galaxies are $\sim 5\times 10^{-11} {\rm yr}^{-1}$.
The four low mass bulges have high specific star formation rate are star 
forming bulge, although their low mass also increase the specific star formation.
These 10 bulges span from passive bulge ($\log$ sSFR $<-10.5$ Gyr$^{-1}$) to active
bulge ($\log$ sSFR $<-10.0$ Gyr$^{-1}$), the threshold of sSFR is suggested
by \citet{fisher10}. The specific star formation rates of the bulges are independent of their
ellipticities. The two plots in Fig.~\ref{fig:pseudo} illustrate the diversity of bulges in the
NIHAO sample.

\subsection{Convergence radius}

The minimum radius above which the results of a simulation are not
significantly affected by the finite resolution is crucial in the numerical study
of bulges. Although running a higher resolution simulations is a way to determining the numerical convergence, it would be extremely expensive for a simulation suite. A common criterion for convergence has been suggested by
\citet{power03} for collisionless simulations, based on the two body
relaxation time-scale for particles in a gravitational potential. 
This criterion ensures that the mean density inside
the convergence radius is within 10\% of the converged value 
obtained in a simulation of much higher resolution.

\citet{tollet15} study the cusp or core of dark matter density 
profiles with the NIHAO suite. They find that this convergence radius, normalized to virial radius, is quite 
constant for all NIHAO galaxies, as all NIHAO galaxies have similar
numbers of particles. At $z=0$, this convergence radius is $0.4 - 0.7$ percent of $\Rvir$ which is comparable with the mean radial
distance of the bulge star particles at $z = 0$. However, as noticed by
\citet{schaller15}, this convertence criterion is too restrictive as it applies to pure DM simulations. As the magnitude of the 
differences between the profiles from hydrodynamical simulations and 
pure DM simulations are significantly larger than 10 percent,
the criterion can be relaxed to the 20 percent level. 
Since we study the stars in the bulge region, where star particles dominate, the convergence radii should
be smaller than in \citet{tollet15}.
Following the prescription of \citet{schaller15}, the convergence radius 
for the sample in this work derived by star particles is of 
the order of 0.08 - 0.1 percent of $\Rvir$ ($\sim$ 100 pc) and the typical
bulge size is in the order of 1 kpc, i.e.\ about an order of magnitude larger, so that the majority of
bulge star particles are comfortably above this resolution limit. 
The range of the convergence radii is shown in the Fig.~\ref{fig:dist_loc_vel}
as grey bar, compared with the radial distribution of bulge stellar
particles at redshift $z = 0$.
Also note that the convergence radii decrease as the redshift increases.
Hence, we expect the migration processes of bulge stars not to be significantly 
impacted by numerical resolution effects.

\subsection{Tracking bulge particles across time}

SPH simulations have the advantage that particles can be easily tracked across time. 
Therefore, we can determine the dynamic history of each stellar particle that 
ends in the bulge at $z=0$. 
Each particle has one specific identifier (id),
which can link every dark matter, gas and star particle to their progenitors
through the whole simulation. Each gas particle may form several (up to 3) star particles. When a new star particle forms from a gas particle, a linker id is created to keep track of the gas particle that produced this particular star particle.

Due to the relatively large time interval of $\sim 200$ Myr between snapshots, we cannot determine the 
exact location and velocity when the star particles form from gas particles.
When a star particle first appears, it has already moved for a certain time ($0-200$~Myr).
Therefore, we also keep track of the location and velocity of the corresponding star-forming gas particle in the previous snapshot. This gas particle, that will have formed a new star particle in the immediate next snapshot is here called a ``parent gas particle''. It is difficult to interpolate from a parent gas particle to a new star particle given that the time interval of $\sim 200$ Myr is comparable to a dynamical time. In future work, we aim to overcome this limitation through the use of 8-times more outputs (simulations in progress).

\section{Formation of bulge} \label{sec:formation}

\subsection{Original location of bulge stars}

\begin{figure*}
	\includegraphics[width=\textwidth]{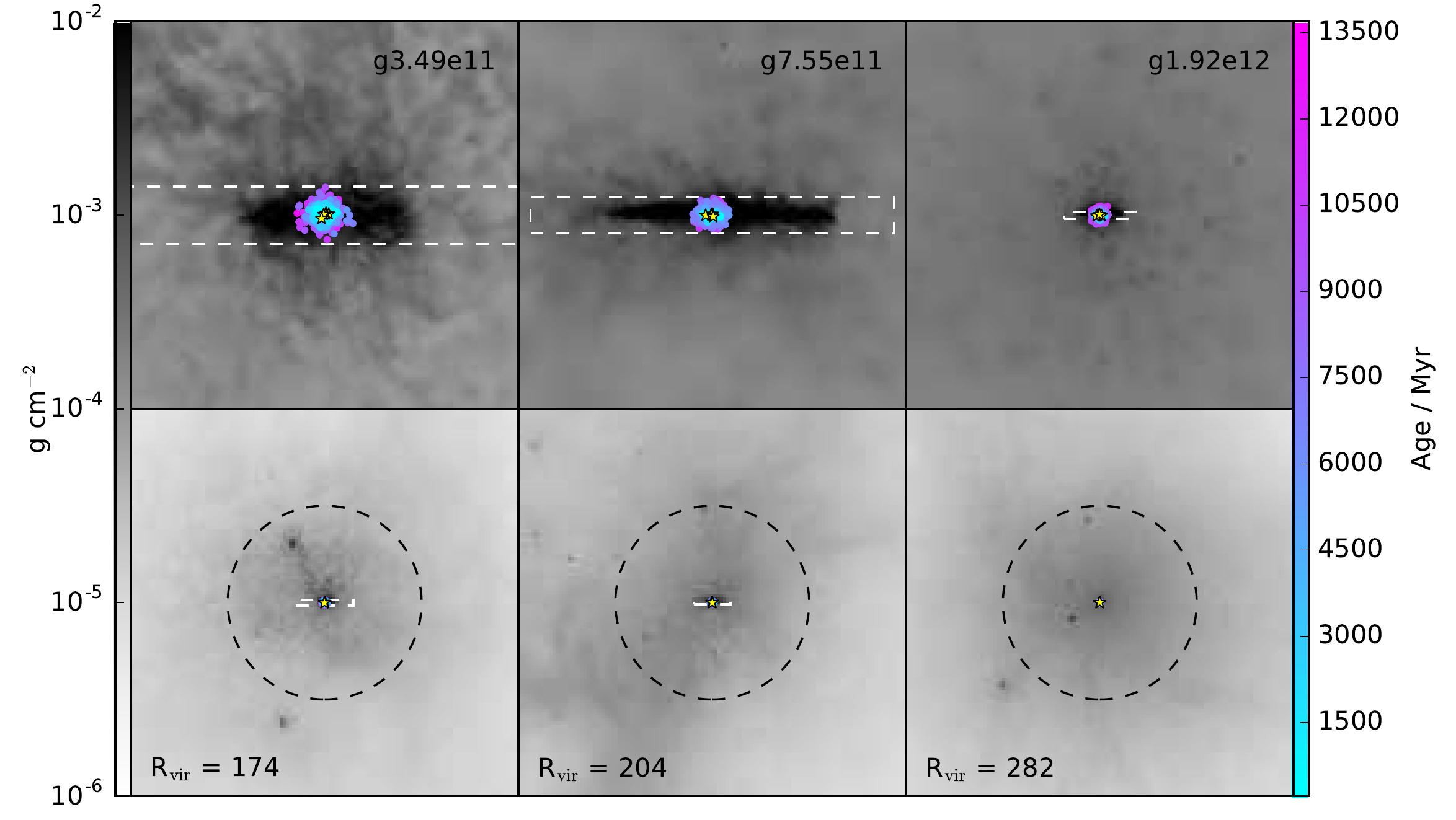}
    \caption{Column density map of gas at z = 0. The size of upper panels is 40\%
         of virial radius at one side and the size of lower panels is 
         10 times larger than the upper panels. The dashed circles are virial
         radius and the dashed rectangles are defined as galaxy region.
         The points are star particles colour coded by their ages and the
         yellow stars are newly formed star particles which formed between
         last two snapshots of the simulations.
         \label{fig:gasmap}}
\end{figure*}

\begin{figure}
	\includegraphics[width=\columnwidth]{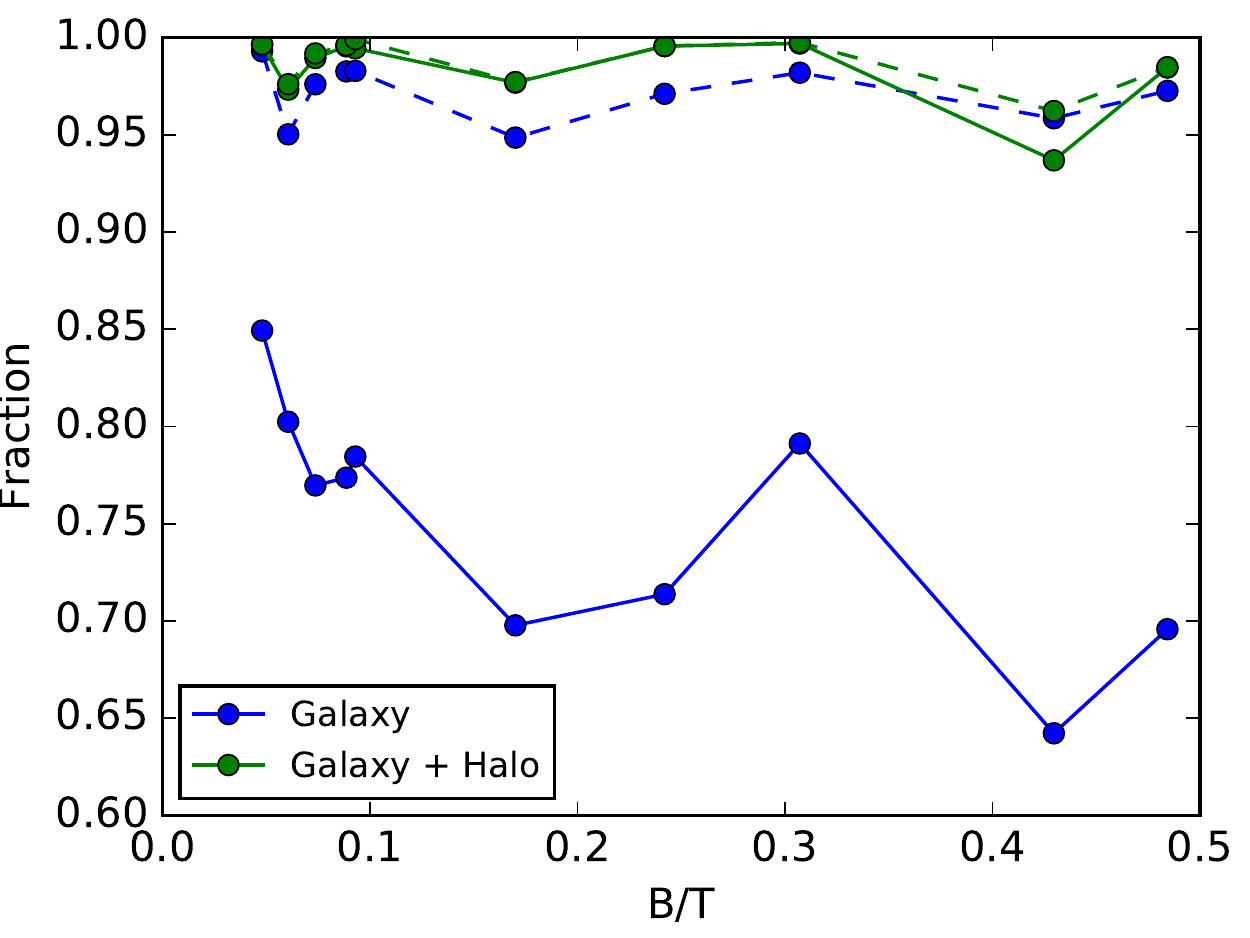}
    \caption{Accumulated mass fraction of bulge stars formed in galaxy region 
         (blue) and in virial radius (green).
         Solid lines show the fraction calculated by gas parent 
         particles and dashed lines show the fraction calculated by
         newly formed star particles.
	 Green, dashed line is nearly identical to the green, solid line.
         \label{fig:origin}}
\end{figure}

Let us now explore where the bulge stars at $z = 0$ formed at higher redshifts in our sample of 10 simulated late-type galaxies. We distinguish three regions of origin, shown in Fig.~\ref{fig:gasmap}: (1) the `galaxy', defined by a cylinder (white dashed rectangle) of
radius $R_{\rm cyn} = 5r_{\rm e}$ and height $h_{\rm cyn} = r_{\rm e}$, where $r_{\rm e}$ is the redshift-dependent half-mass radius of galaxy; (2) the `halo', defined as the region outside the galaxy, but inside the virial radius (black dashed circle) at the given redshift; (3) the `environment', defined as the region outside the halo. Processes in the `halo' or the `environment' are often called {\it ex situ} processes in the literature. In principle, it is possible to subdivide the galaxy region in the aim of separating stars formed in a central star-burst from stars formed in a disc. However, we found this division to be cumbersome at higher redshift, especially at cosmic times when the disc is hardly stable and thick (high pressure support), which are indeed the cosmic times where most stars formed. We therefore avoid labelling stars as bulge or disc stars when they form, but instead quantify the star-formation dynamics inside the galaxy by tackling the radius, velocity and AM evolution of every star that lands in the bulge at $z=0$.

The fractions of bulge stars formed in each region are shown in Fig.~\ref{fig:origin} as a function of the final $B/T$ ratio. These fractions rely on the location of the parent gas particles (solid line) and newly formed star particles (dashed line) in the following snapshot (i.e.$\sim 200$ Myrs later), since we have no information about the exact location of the star formation in between snapshots. The fractions calculated by the parent gas particles show that the majority (about 75\%) of them have already settled in the galaxy region when they form stars. Most of the remaining bulge stars (20\% of all bulge stars) were formed in the halo, leaving no more than a few percent bulge stars form outside the halo. In other words, mergers do not play an important role in the bulge formation in our sample.

When measuring these fractions using the newly formed stars instead of the parent gas particles, the picture looks somewhat different. We still find that most of the bulge stars ($>95\%$) form inside the virial radius. However, the balance between halo and galaxy is different: even more bulge stars (95\% on average) now appear to originate from the galaxy. This difference suggests that most parent gas particles inside the halo convert into stars as soon as they reach the galactic disc -- an explanation, which is in line with the higher density and thus lesser stability and increase self-shielding expected in the galaxy. 

\subsection{Evolution of AM of bulge stars} \label{sub:ang}

\begin{figure*}
    \includegraphics[width=0.75\textwidth]{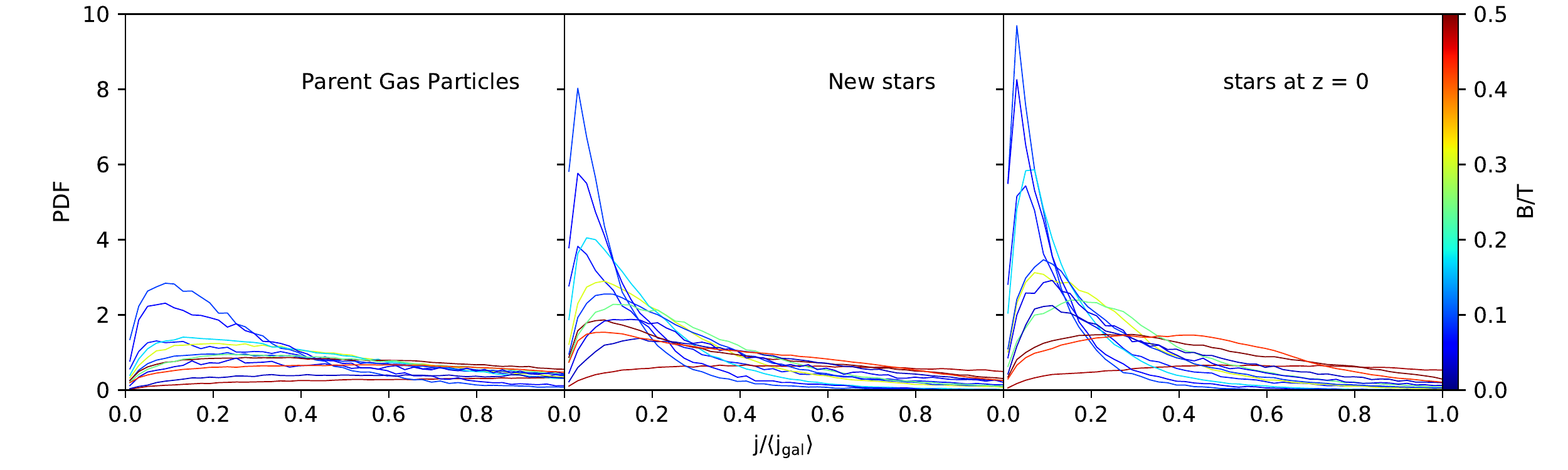}
    \includegraphics[width=0.24\textwidth]{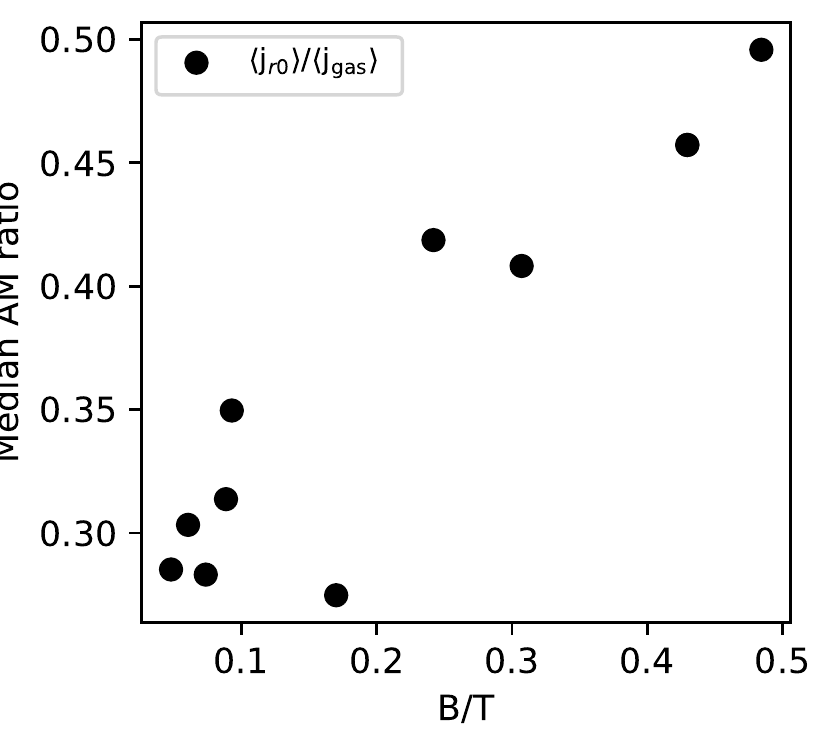}
    \caption{Left: Normalized distribution of specific AM of parent gas particles
         (left), newly formed bulge star particles (middle) and bulge stellar 
         particles at z = 0 (right). 
         The curves are colour coded by the bulge-to-total ratio.
         Right: The median of specific angular
         momentum ratio between bulge star particles at z = 0 $\langle
         j_{r0} \rangle$ and parent gas particles $\langle j_{gas} \rangle$.
         \label{fig:amhist}}
\end{figure*}

The stars in bulges are always the low AM component of the galaxies, even in the case of rotationally supported bulges, due to their small radius relative to the disc.
As discussed before, most of the bulge stars (95\%) of the galaxies in our simulation formed inside the galaxy. One scenario for this bulge formation in the galaxy is that cold gas in the disc forms stars with high AM and the stars then lose their AM by dynamical friction to migrate into the bulge region.
The other possibility is that the cold gas first dissipates its AM and then forms
stars with low AM in or close to the central region. The probability distribution function of the specific AM $j$ can be used to clarify the scenario
and provide useful insights regarding the details of bulge formation.

The left panel of Fig.~\ref{fig:amhist} shows the distribution of $j$ of the parent gas particles (left), newly formed bulge star particles (middle) and 
bulge star particles at $z = 0$ (right) for each simulation. 
The specific AM is normalized by the mean specific AM $\langle j_{gal} \rangle$ of all particles (stars, gas and dark 
matter) within 20\% of the virial radius at $z = 0$.
Each solid 
curve represents a galaxy and the curves are coloured by their $B/T$ value. 
Qualitatively, the curves with different $B/T$ have significantly different
shapes and evolution features.
For the galaxies with higher $B/T$, the initial distributions of parent gas
particles have a lower peak and longer tails compared to the galaxies with
lower bulge fractions. 
For newly formed bulge stars and $z = 0$ bulge stars, the peaks of the 
distributions move inwards, which means that the stars
lose their AM during the processes of forming the bulge.
However, the distributions of galaxies with lower $B/T$ have much smaller widths
and higher peaks than the galaxies with higher $B/T$. A similar trend is found in observations of spiral galaxies \citep{sweet18}.

The right panel of Fig.~\ref{fig:amhist} shows the
median of the specific AM ratio between bulge star particles at $z = 0$ and
parent gas particles. A ratio of unity would mean that $j$ of bulge stars is conserved from (just before) their formation to $z=0$, while smaller ratios indicate a loss of $j$.
The plot shows a clear trend for a higher loss in $j$ in the case of smaller bulge final fractions $B/T$. The specific AM ratio between bulge star particles at $z=0$ and newly formed stars (rather than star-forming gas) shows a similar trend with the bulge fraction $B/T$, but the mean ratio is higher ($\sim 80\%$).
In the following section, the implications of the loss in $j$ on the stellar dynamics will be discussed in more detail.

\subsection{Evolution of locations and velocities of bulge stars} \label{sub:dis}

According to the definition of the specific AM $j$, a variation in $j$ of the bulge stars can come from a variation of the orbital radius or the orbital velocity.

\begin{figure*}
    \includegraphics[width=\textwidth]{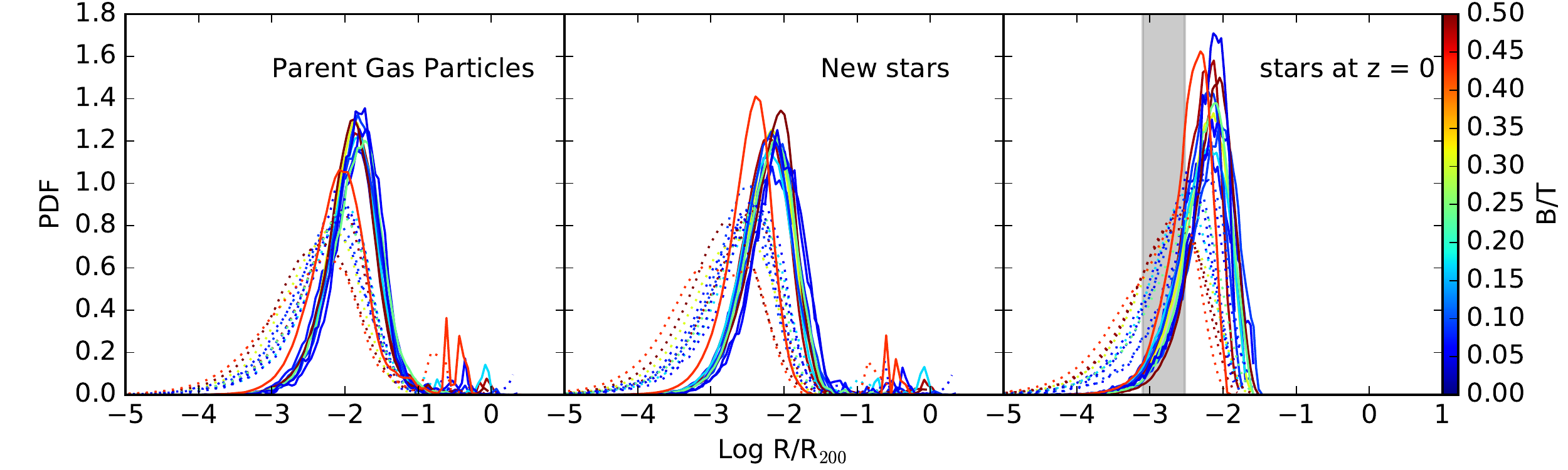}
    \includegraphics[width=\textwidth]{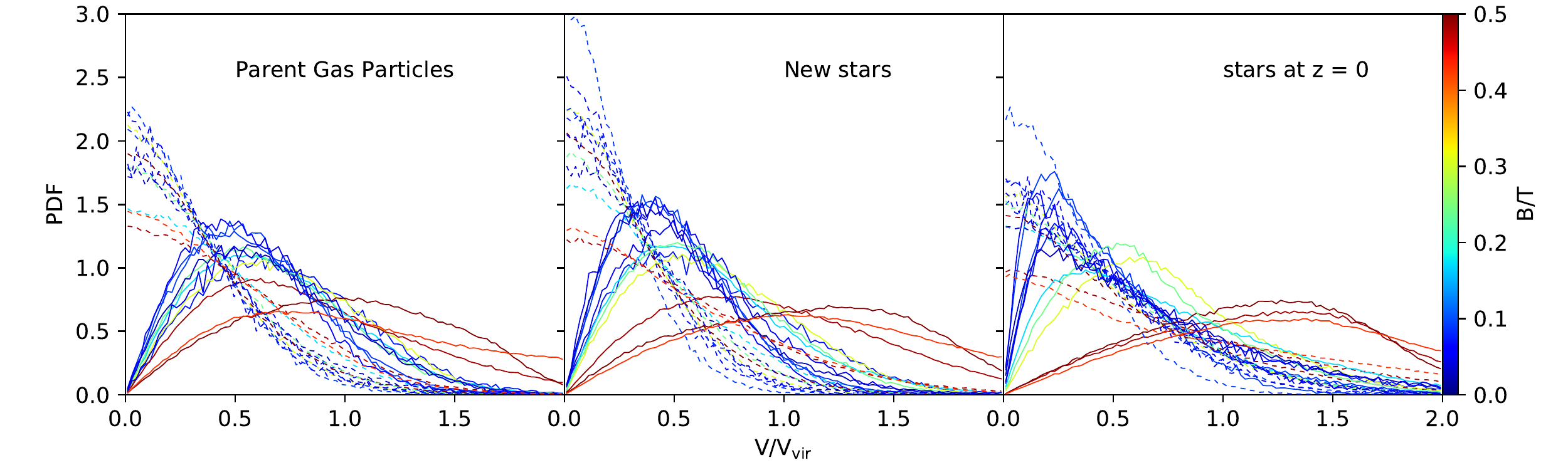}
    \caption{Upper panel: Normalized distribution of radial distance (solid line) and height
         (dotted line) with fixed z-direction. The left panel is for the parent gas
         particles, the middle panel is
         for the newly formed  bulge stars and the right panel is for the bulge stars
         at $z = 0$. The grey vertical region shows the range of convergence radii.
         Lower panel: Normalized distribution of azimuthal velocity (solid line) and 
         z component of velocity 
         (dotted line) with fixed z-direction. The left panel is for the parent gas
         particles, the middle panel is
         for the newly formed  bulge stars and the right panel is for the bulge stars
         at $z = 0$.
         The curves are color coded by the $B/T$ ratio.
         \label{fig:dist_loc_vel}}
\end{figure*}

\begin{figure}
	\includegraphics[width=\columnwidth]{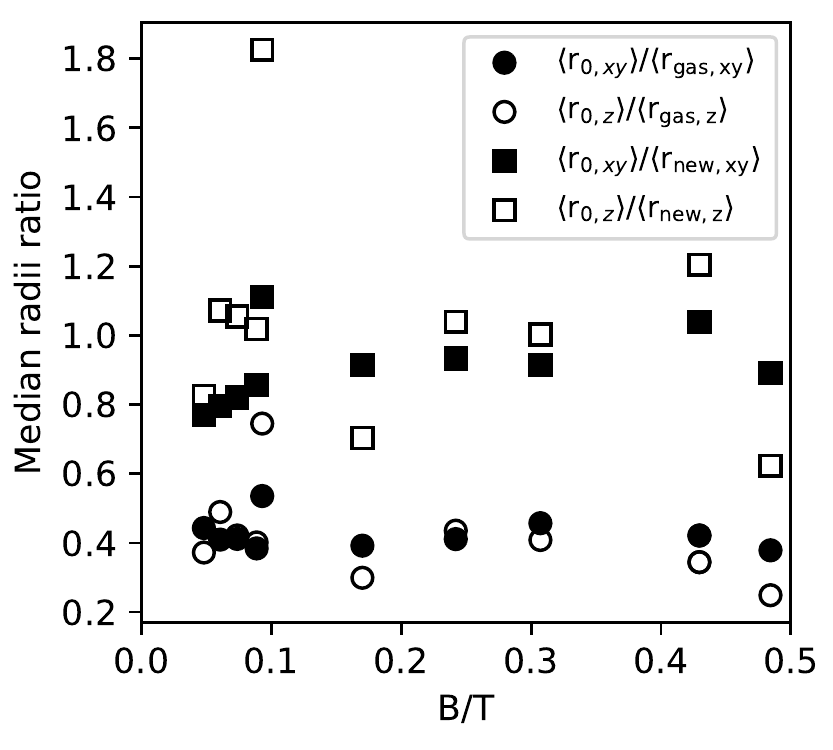}
    \includegraphics[width=\columnwidth]{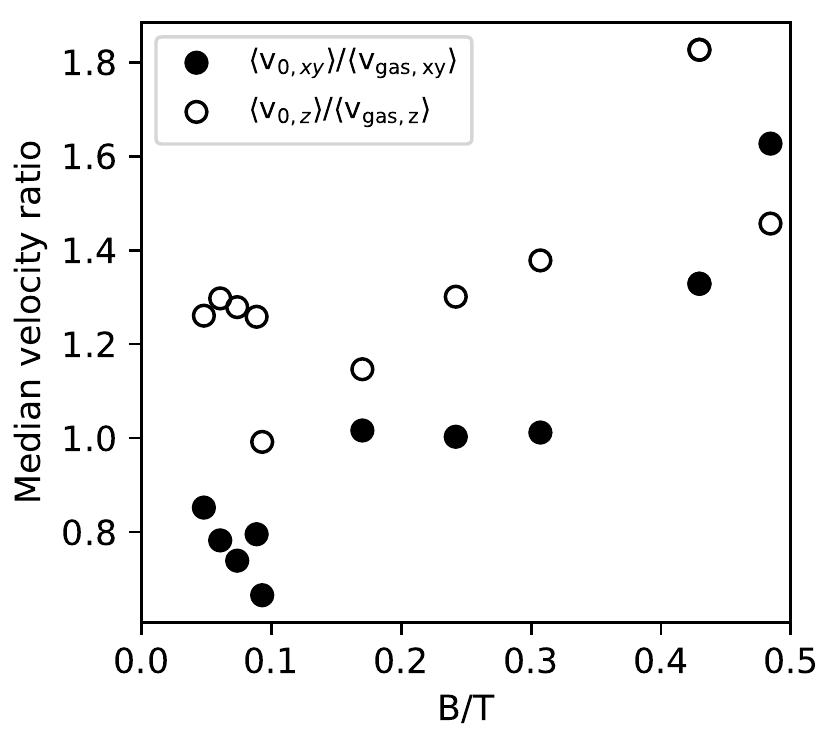}
    \caption{Upper panel: median ratio of radial distance and height between parent gas particles and bulge star particles at $z = 0$. The squares show the median ratio between newly formed star particles and bulge star particles at $z=0$.
         Lower panel: median ratio of azimuthal velocity (black points) and vertical velocity (white points) between parent gas particles and bulge star particles at $z = 0$.}
         \label{fig:m_dis_vel}
\end{figure}

We compute the distributions of the orbital radii and velocities of the parent 
gas particles, newly formed star particles and bulge star particles 
at $z = 0$, in analogy to the $j$-distributions in Fig.~\ref{fig:amhist}.
In Fig.~\ref{fig:dist_loc_vel}, the upper panel shows the normalized distributions 
of the radial distance and the vertical distance, in a coordinate system, whose orientation is fixed to that of the final snapshot at $z=0$.
The distances are normalized to the virial radius at $z = 0$.
The general trend is that both the radial and vertical distance decrease with time, i.e.~the  bulge stars migrate inwards.
In contrast to the $j$-distributions, these distance-distributions have an almost universal shape with no clear trend as a function of $B/T$. Two galaxies (red and light blue lines in the upper panel of Fig.~\ref{fig:dist_loc_vel}) have a substantial fraction bulge stars formed outside the virial
radius and there is no statistically significant relation with the galaxies' bulge-to-total ratio.
It is necessary to note that the vertical distance of most parent
gas particles is smaller than their radial distance shown by the distributions.
This means that the migrations of the bugle particles are not isotropic, and they are preferentially 
near the disc plane during the migration.


The lower panel in Fig.~\ref{fig:dist_loc_vel} shows the normalized distribution of the
azimuthal and vertical velocity. The distributions and their variations have a clear dependence on  
the galaxies' morphology.
The distributions of azimuthal velocities (solid lines) and their variations for the galaxies 
with lower $B/T$ ($<10\%$) are similar to the $j$-distributions show in Fig.~\ref{fig:amhist} (a). In particular, the mean velocities decrease with time and the width of the distributions gets narrower.
At higher $B/T$ ratios, the distributions of azimuthal velocities show a different trend.
The peaks of the distributions for the parent gas particles are lower
and the shapes of the distributions are wider
than the distributions of galaxies with lower $B/T$. 
The particle velocities generally decrease in the panels for the newly formed star particles and bulge star particles
at $z = 0$. Fig.~\ref{fig:m_dis_vel} highlights the main trends of Fig.~\ref{fig:dist_loc_vel}.
Interestingly, as long as $B/T$ $>20\%$, the median of velocity of the bulge stars actually increases from the time these stars form to $z=0$.
their z-component of the velocity is negligible comparing with their azimuthal
velocity.

\section{Discussion} \label{sec:discuss}

\subsection{Linking the evolution of specific angular momentum, velocity and distance} \label{sec:link}
\label{sec:linkjrv}

\begin{figure}
	\includegraphics[width=\columnwidth]{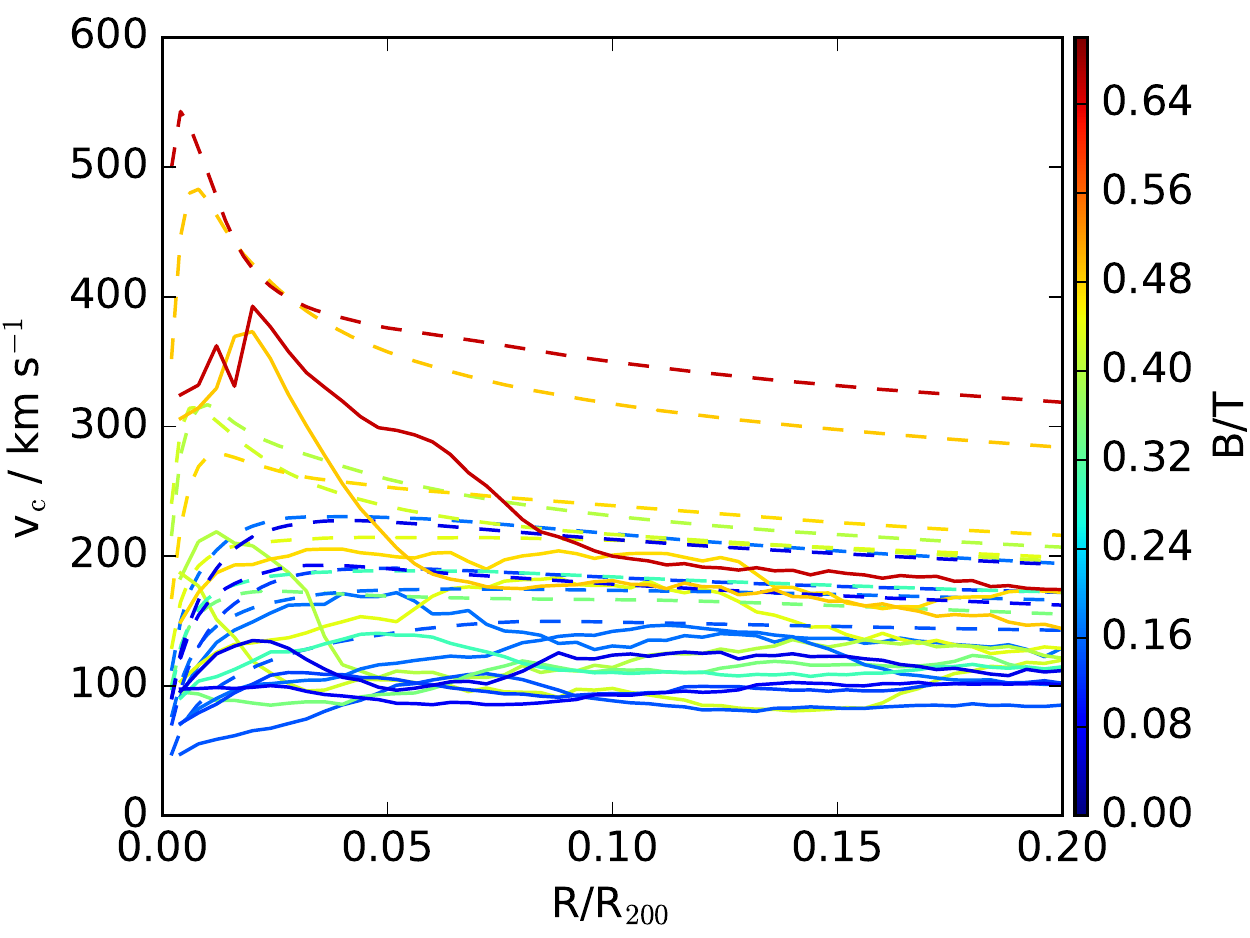}
    \caption{Rotation curves of the median rotation velocity (solid line) and 
         spherically circular velocity $\sqrt{GM/r}$ (dashed line) colour-coded with the $B/T$ ratio.
         \label{fig:rot_cur}}
\end{figure}

In the left panel of Fig.~\ref{fig:amhist} we tackled the evolution of bulge star particles from their parent gas particle phase to the final stage at $z=0$.
The specific AM mainly decreases in the parent gas phase, immediately ($<200$ Myrs) before the stars form: as the cooling gas reaches the disc region it dissipates the AM by fluid viscosity and dynamical friction. Once the stars have formed, they are no longer subject to viscosity, and hence the AM loss continues at a much lower rate, set purely by gravitational torques. Globally, we find that more AM is dissipated for smaller final bulge fractions, as shown in the right panel of Fig.~\ref{fig:amhist}.

Given this loss of AM, we can then attempt to explain the change in the radius and velocity distributions (Fig~\ref{fig:dist_loc_vel} and Fig~\ref{fig:m_dis_vel}). In a Keplerian framework, the circular velocity at a given radius is set by the enclosed mass, which is
$v=\sqrt{GM\left(r\right)/r}$
for spherical mass distributions. 
Hence the balance between change in radius and velocity depends on this mass, which itself depends on the bulge mass. For small bulge masses ($B/T$ $<0.1$), the specific AM reduction is paralleled by a decrease in radius \emph{and} velocity. However, for higher bulge masses, this AM reduction is achieved by a loss in radius under increase in velocity. The circular velocity simply has to increase due to the deeper potential caused by the bulge itself. This is apparent in Fig.~\ref{fig:rot_cur}, which shows the rotation curves, measured by the median circular velocity (solid lines) and the enclosed mass (dashed lines) for different $B/T$ ratios. The rotation curves exhibit qualitatively different shapes for small and large $B/T$, with larger values leading to more peaked rotation curves.

\subsection{Bulge formation by migration?}

\begin{figure}
	\includegraphics[width=\columnwidth]{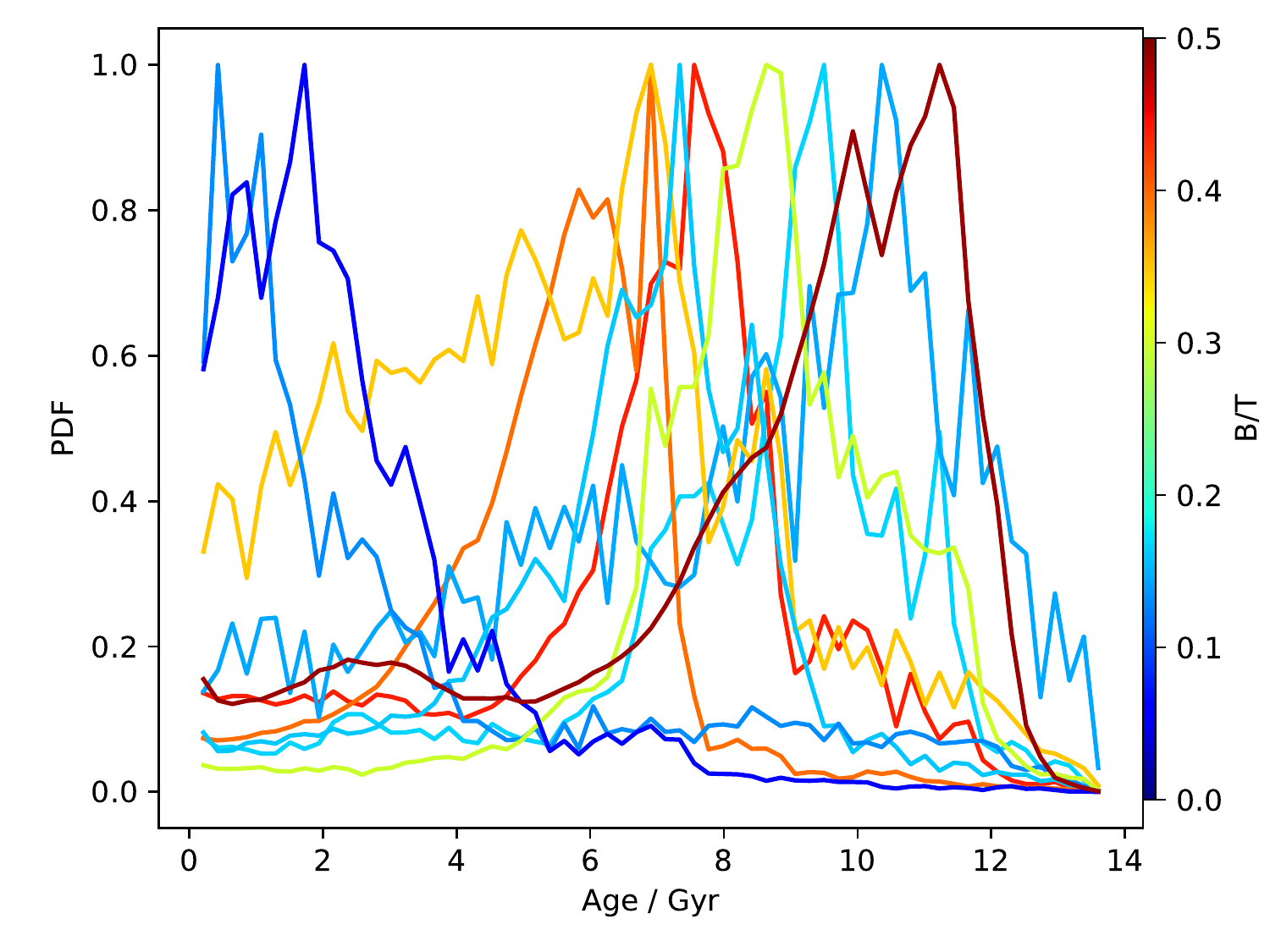}
    \caption{Normalized age distribution of the bulge star particles for the NIHAO galaxies analysed in this paper. The colour of the lines indicates the B/T ratio, as labelled by the colourbar.
         \label{fig:age_dist}}
\end{figure}

The global star formation density in the Universe peaks around $z=1\sim 3$ \citep{hopkins06}.
In this epoch star-forming galaxies were more ``clumpy'' than in the local Universe \citep{abraham96, conselice00,
elmegreen05}.
The migration of bulge stars towards the centre is normally discussed at
the level of clumps. 
One scenario argues clumps have lifetimes long enough to allow them to migrate 
inwards forming the galaxy bulge \citep{bournaud14,forbes14}. This would predict that clumps closer to the galaxy centre are older, which is indeed suggested by observations (Lamberti et al., 2018, submitted). Alternatively, older clumps may be closer to the galaxy centre as a result of inside-out formation of the disc \citep{murray10,genel12}. Fig.~\ref{fig:age_dist} shows the normalized age distributions of the bulge star particles of each galaxy at $z=0$. Most bulges are dominated by old stars (formed at $z > 1$), and only two galaxies with very small bulge-to-total mass ratios, host young bulges. \citet{guedes13} and \citet{okamoto13} found similar results; that is, the bulk of the pseudo-bulge stars forms quickly at high redshift by a combination of non-axisymmetric disc instabilities and tidal interactions or mergers. They concluded that the main formation mechanism of pseudo-bulges are high-redshift starbursts, rather than secular evolution of galactic discs.

To investigate the migration of bulge stars, we here took a different approach: we do not explicitly consider clumps or distinguish between star formation in the disc and bulge, but instead focus directly on the change in radius of the population of bulge stars at $z=0$. 
The reason is that the resolution of the simulations is insufficient to define the clumpy structure. This is because clumps have shown to have sizes from dozens of pc to a kpc \citep{livermore15}. We found that the majority of the bulge is contributed by stars formed in the ``galaxy'', that is in the disc+bulge region, mainly around less than 1\% $\Rvir$. By comparing the distribution of radial and vertical distances (solid and dotted lines in Fig. 8, top), we find that most of this star formation took place in an oblate region, akin to a thick disc. From there, the radii of the stellar orbits decrease, on average, by a factor 2.5, for all $B/T$ values. While the present simulations do not have the time-resolution to tackle spiralling orbits of clumps, the global change in radius nonetheless argues for a migration scenario similar to that of \citet{bournaud14}. 
Similar simulations with 8 times more snapshots are in preparation. In the future we plan to work on distinguishing inner and outer disc origin for bulge stars throughout cosmic time and connect this to the AM build-up of the bulge.

\subsection{AM evolution as a function of $B/T$}

\begin{figure}
	\includegraphics[width=0.9\columnwidth]{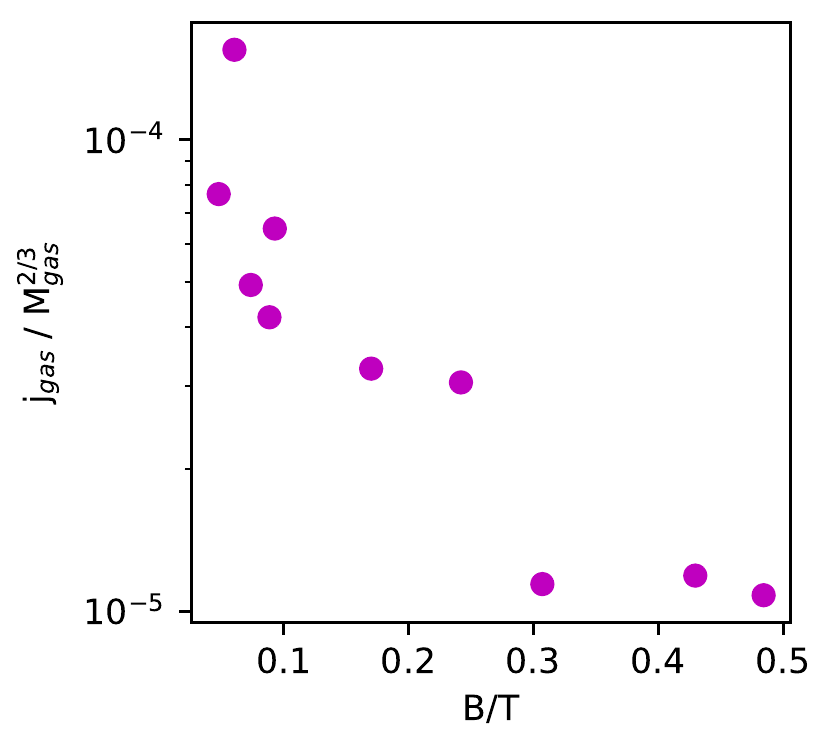}
    \caption{The $j/M^{2/3}$ of the parent gas particles as the function of the bulge fraction $B/T$. 
         \label{fig:bulge_age}}
\end{figure}

In Fig.~\ref{fig:amhist} (right), the AM loss of bulge stars from their formation to $z=0$ appears to be negatively correlated with the final bulge fraction $B/T$. This is an important result, which calls for a physical explanation. A plausible scenario is that both the AM loss and the final bulge fraction correlate with the spin parameter $\lambda \propto j/M^{2/3}$ \citep{peebles69} of the stellar system. Let us consider each of these correlations separately.

Fig.~\ref{fig:b_d} shows that bulges and discs in the simulations approximately reproduce the respective empirical $j$--$M$ relation of field galaxies. Empirically, classical bulges, pseudo-bulges and elliptical galaxies all fall on roughly the same power-law relation of slope $2/3$ \citep{fall18}. In other words, all these systems have a universal stellar $\lambda$. If we take this empirical fact for granted, bulge stars that form with a higher $\lambda$ value are thus bound to loose more AM. Hence $\lambda$ and the AM loss are positively correlated. Recently, \citet{lagos18b} analysed a large sample of simulated galaxies by combining the EAGLE and Hydrangea (cluster) simulations (\citealt{schaye15, bahe17}). They explicitly explored the relation between the stellar and halo spin parameters, and found a positive correlation. Galaxy mergers tend to wash out that correlation, which means that galaxies in the absence of mergers follow the halo $j$ closely (see also \citealt{zavala16}).

On the other hand, Fig.~\ref{fig:bulge_age} shows the relation between $j/M^{2/3}$ of all the parent gas particles and the bulge fraction $B/T$. Clearly, gas with higher spin parameter forms a smaller bulge, similarly to empirical models which find that galaxies with a lower stellar or baryonic $j/M^{2/3}$ values exhibit more significant bulges \citep{romanowsky12,obreschkow14,sweet18,fall18,posti18}. While this relation is theoretically not yet fully understood, it is likely a consequence of global (bar-mode) or local (Toomre) instabilities scaling roughly as $j/M^{2/3}$ \citep{obreschkow14}. 

Having established that higher $\lambda$ values (of star-forming gas or newly formed stars) lead to less bulge formation as well as to more AM loss when forming a bulge, means that the AM loss of bulge stars must be negatively correlated with the final bulge mass fraction. In other words, the AM ratio between the parent gas of bulge stars and bulge stars at $z=0$ correlates \textit{positively} with the final bulge fraction $B/T$, explaining Fig.~\ref{fig:amhist} (right). 

\section{Conclusions} \label{sec:summary}

In this paper, we analysed the dynamic evolution of stars that end up in the central bulge of disc-dominated galaxies at $z=0$, using a sample of 10 galaxies from the NIHAO simulation. The `bulge' is defined as all excess mass above an exponential disc in the central region of galaxies, dominated by rotation-supported pseudo-bulges. The simulated galaxies lie on a similar $j_{\rm bar}$--$M_{\rm bar}$ relation as observed star-forming spiral galaxies in the local universe. Furthermore, the galaxies with large $B/T$ ratios have lower $j_{\rm bar}$ values, also in relatively good agreement with modern (interferometric) observations. These observational verifications justify the use this sample of simulated galaxies to investigate the formation and AM evolution of bulge stars. Our results can be summarized as follows.

\begin{itemize}
  \item By tracking bulge star particles back in time to their parent gas particles, we found that $\gtrsim 95\%$ of the stellar mass in bulges at $z = 0$ formed {\it in-situ} in the galaxy, while about $\sim 3\%$ of the mass stems from star formation outside the galaxy but inside the virial radius (i.e.~in satellites). A small and comparable mass fraction of the bulge stars formed outside of the virial radius in the simulated sample.
  \item The AM evolution of the bulge stars, from their formation to $z=0$, shows a clear correlation with the final $B/T$ value of the galaxy: the lower the final $B/T$ value the more AM is dissipated away from the bulge stars (Fig.~8, right).
  \item On average, bulge stars move towards the galaxy centre by about a factor 2, \emph{irrespective} of the final $B/T$ ratio. By contrast, the circular velocity of the bulge stars tends to slightly increase with time, and this increase scales with the final $B/T$ value.
 \end{itemize}

Specific AM $j$, mass $M$ and bulge-to-total ratio $B/T$ obey a strong correlation in regular galaxies \citep{romanowsky12}, even within the subset of gas-rich star-forming spirals \citep{obreschkow14}. The bulges decomposed photometrically means the central overdensity components, this raises the question of how and whether this bulge mass is driven by the AM in the galaxy. Our analysis showed that the AM history of bulge stars in this definition correlates with the galaxy's final $B/T$. Qualitatively, galaxies with higher AM (at fixed mass) and hence a requirement to dissipate more AM to move stars to the centre, tend to form smaller bulges. The details of this connection remain, however, clouded by the limited spatial and temporal resolution in the simulations. By increasing the number of snapshots for one simulation in NIHAO sample, we will soon have the ability to rebuild the trajectories of bulge stars from their formation to $z=0$ in ten-times more detail.

\section*{Acknowledgements}

We are grateful to Pascal Elahi, Matthieu Schaller and Jiang Chang for helpful 
comments and discussions. We also thanks the anonymous referee for a constructive report that helped improve the clarity of the paper. 
The analysis was performed using the pynbody package (http://pynbody.github.io),
which was written by Andrew Pontzen and Rok Ro{\v s}kar in addition to the authors.
This research was supported by Australia Research Council Discovery Project
160102235.
This research was carried out on the High Performance Computing resources at New York University Abu Dhabi;
on the {\sc theo}  cluster of the Max-Planck-Institut f\"ur Astronomie and on the {\sc hydra}  clusters at the Rechenzentrum in Garching.
CL is funded by a Discovery Early Career Researcher Award (DE150100618) and by the ARC Centre of
Excellence for All Sky Astrophysics in 3 Dimensions (ASTRO 3D).
XK acknowledges the support from NSFC project No.11333008 and 973 program No. 2015CB857003.





\bibliography{nihao_bulge}



\appendix

\section{Parameters}

In this appendix we summarize bulge parameters of all galaxies at redshift $z=0$ 
through this paper.
Table~\ref{tab:parameters} contains the simulation ID, baryonic mass $M_{bar}$, baryonic specific angular
momentum $j_{\rm bar}$, photometric bulge-to-total ratio [$B/T$]$_{\rm p}$, kinematic bulge-to-total ratio [$B/T$]$_{\rm k}$, ellipticity $\epsilon$, $V/\sigma$ and specific star formation
rate $\rm sSFR$.

\begin{table*}
\caption{Bulge parameters for galaxies at redshift $z=0$ }  
\label{tab:parameters}
\begin{center}
\begin{tabular}{lllrrllr}
\hline
Simulation ID & $M_{\rm bar}$ & $j_{\rm bar}$ & [$B/T$]$_{\rm p}$ & [$B/T$]$_{\rm k}$ & $\epsilon$ & $V/\sigma$ & sSFR \\
              & $\rm M_\odot$     & kpc km s$^{-1}$ &        &        &        &            & yr$^{-1}$ \\
\hline
	g7.08e11 & $5.22\times10^{10}$ & $9.52\times10^{2}$ & 0.07 & 0.27 & 0.04 & 0.56 & $4.23\times10^{-11}$ \\ 
g7.55e11 & $4.94\times10^{10}$ & $9.11\times10^{2}$ & 0.31 & 0.42 & 0.33 & 0.71 & $2.43\times10^{-11}$ \\ 
g7.66e11 & $6.32\times10^{10}$ & $3.29\times10^{2}$ & 0.48 & 0.46 & 0.29 & 0.86 & $3.09\times10^{-11}$ \\  
g3.49e11 & $1.46\times10^{10}$ & $8.49\times10^{2}$ & 0.09 & 0.42 & 0.05 & 0.61 & $1.47\times10^{-10}$ \\ 
g3.71e11 & $1.64\times10^{10}$ & $2.53\times10^{2}$ & 0.17 & 0.70 & 0.17 & 0.61 & $2.06\times10^{-11}$ \\ 
g5.02e11 & $2.53\times10^{10}$ & $8.03\times10^{2}$ & 0.09 & 0.53 & 0.07 & 0.59 & $2.04\times10^{-11}$ \\ 
g5.36e11 & $2.54\times10^{10}$ & $3.49\times10^{2}$ & 0.06 & 0.76 & 0.07 & 0.60 & $3.42\times10^{-10}$ \\ 
g5.38e11 & $2.82\times10^{10}$ & $6.84\times10^{2}$ & 0.24 & 0.44 & 0.16 & 0.73 & $1.13\times10^{-11}$ \\ 
g6.96e11 & $4.91\times10^{10}$ & $3.78\times10^{2}$ & 0.05 & 0.49 & 0.14 & 0.68 & $2.26\times10^{-10}$ \\ 
g1.92e12 & $1.57\times10^{11}$ & $6.89\times10^{2}$ & 0.43 & 0.27 & 0.40 & 0.82 & $5.21\times10^{-11}$ \\ 

	\hline
\end{tabular}
\end{center}
\end{table*}

\section{Stellar surface density profiles for individual galaxies}

In this Appendix, we present the circularity distribution of total star particles and bulge components decomposed by kinematic and photometric methods, stellar surface density profiles, the exponential
fit for disc component, and synthetic images of the galaxies.

\begin{figure*}
\includegraphics[width=0.4\textwidth]{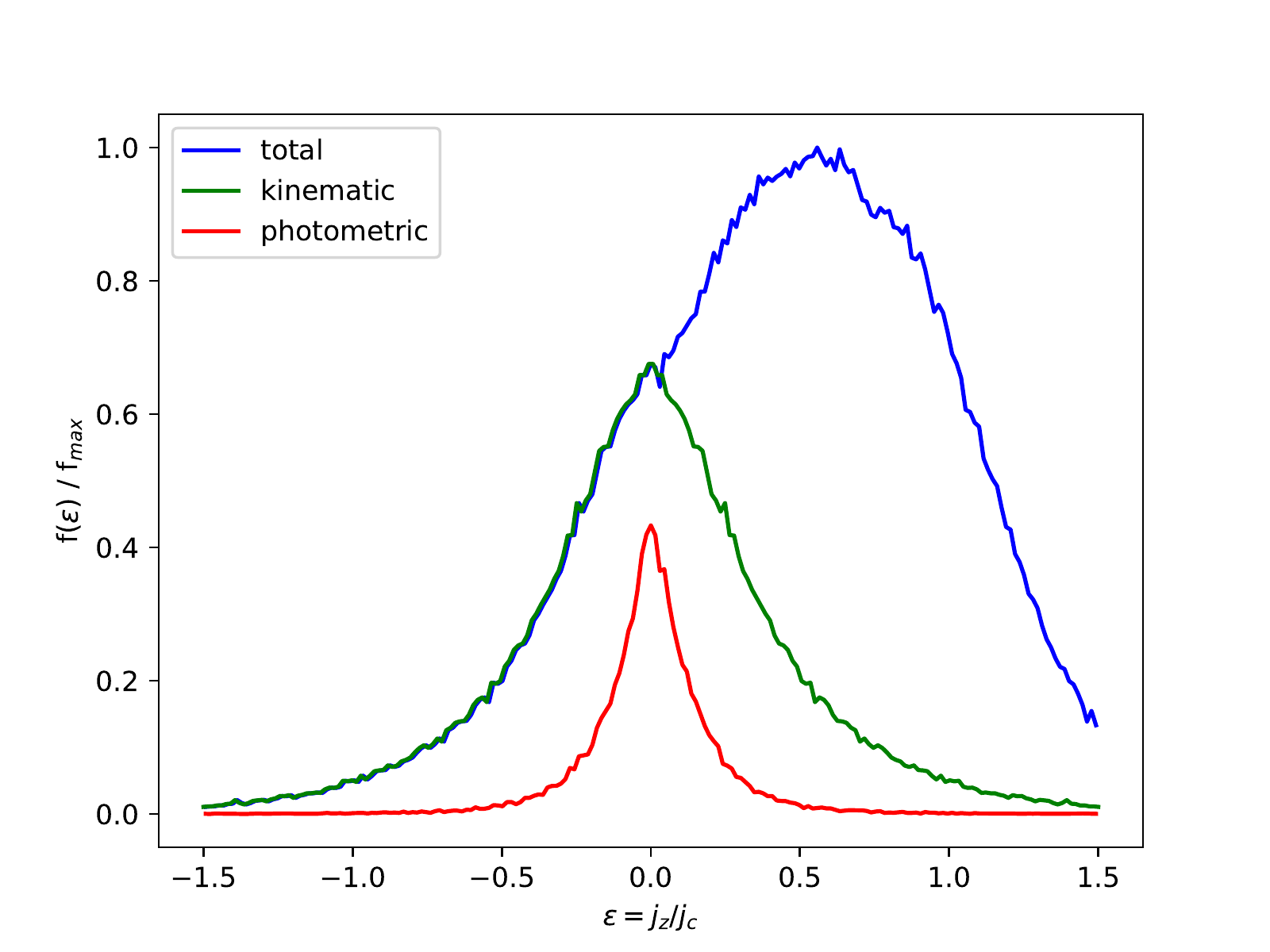}
\includegraphics[width=0.58\textwidth]{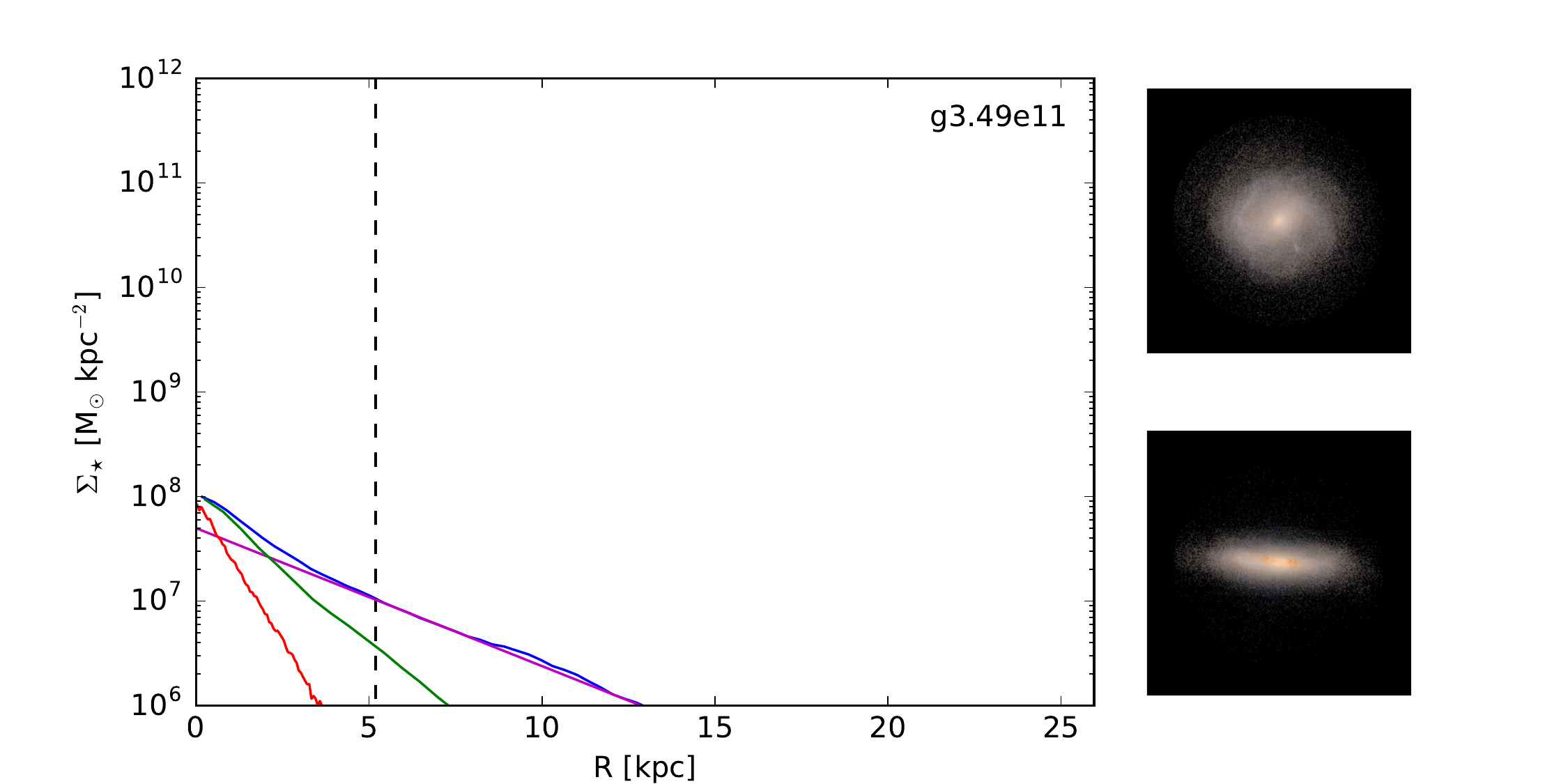}
\includegraphics[width=0.4\textwidth]{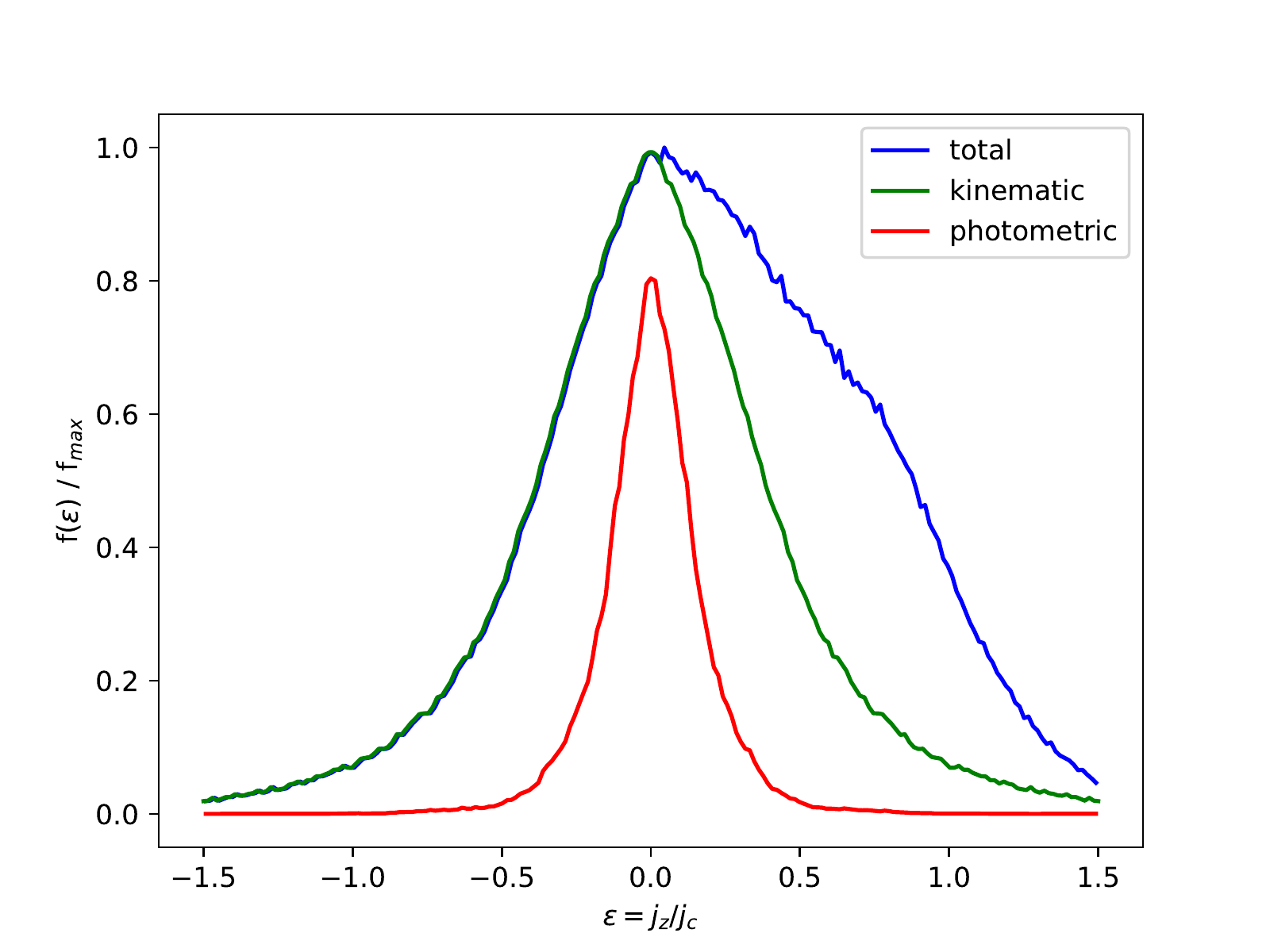}
\includegraphics[width=0.58\textwidth]{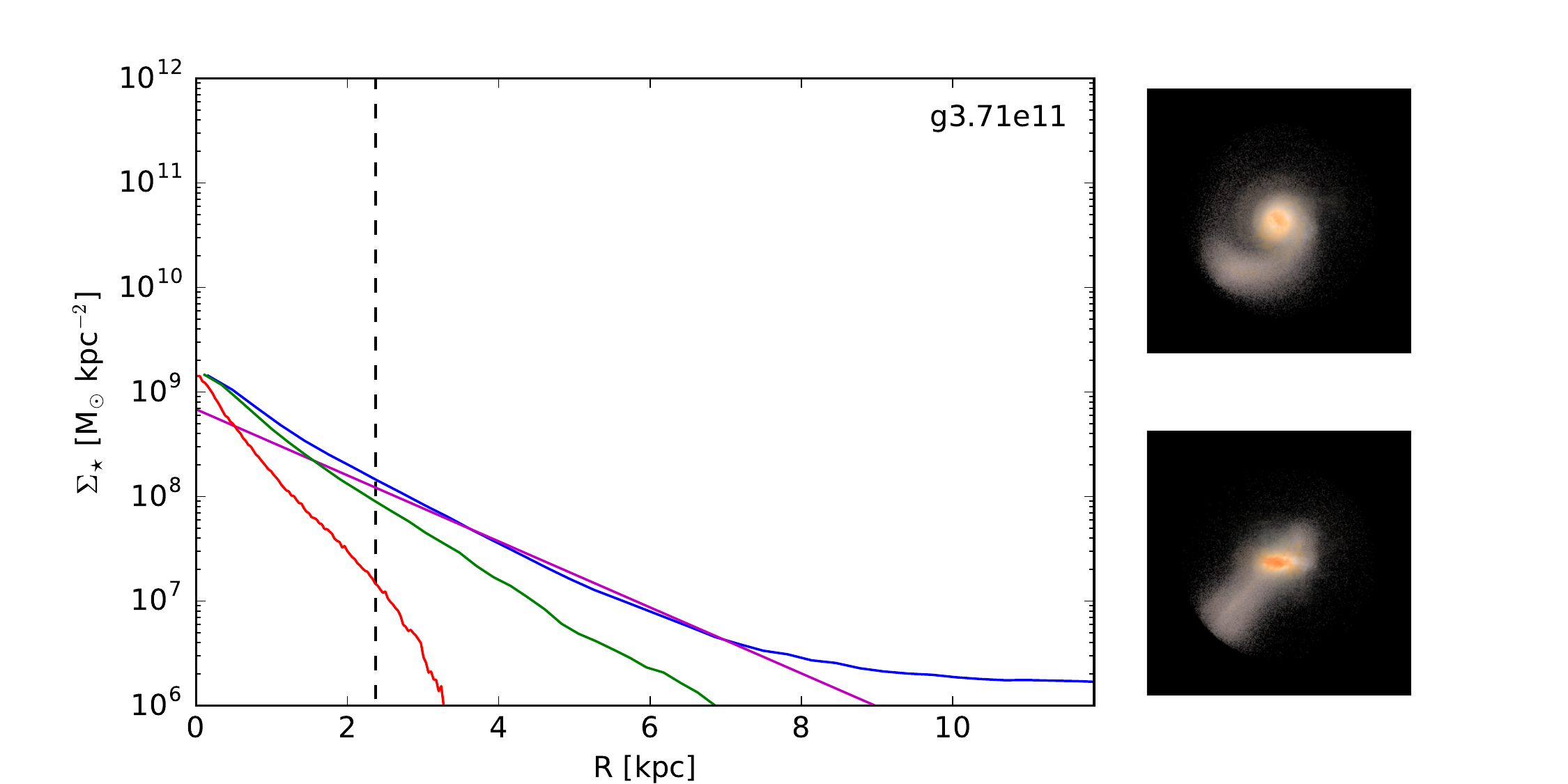}
\includegraphics[width=0.4\textwidth]{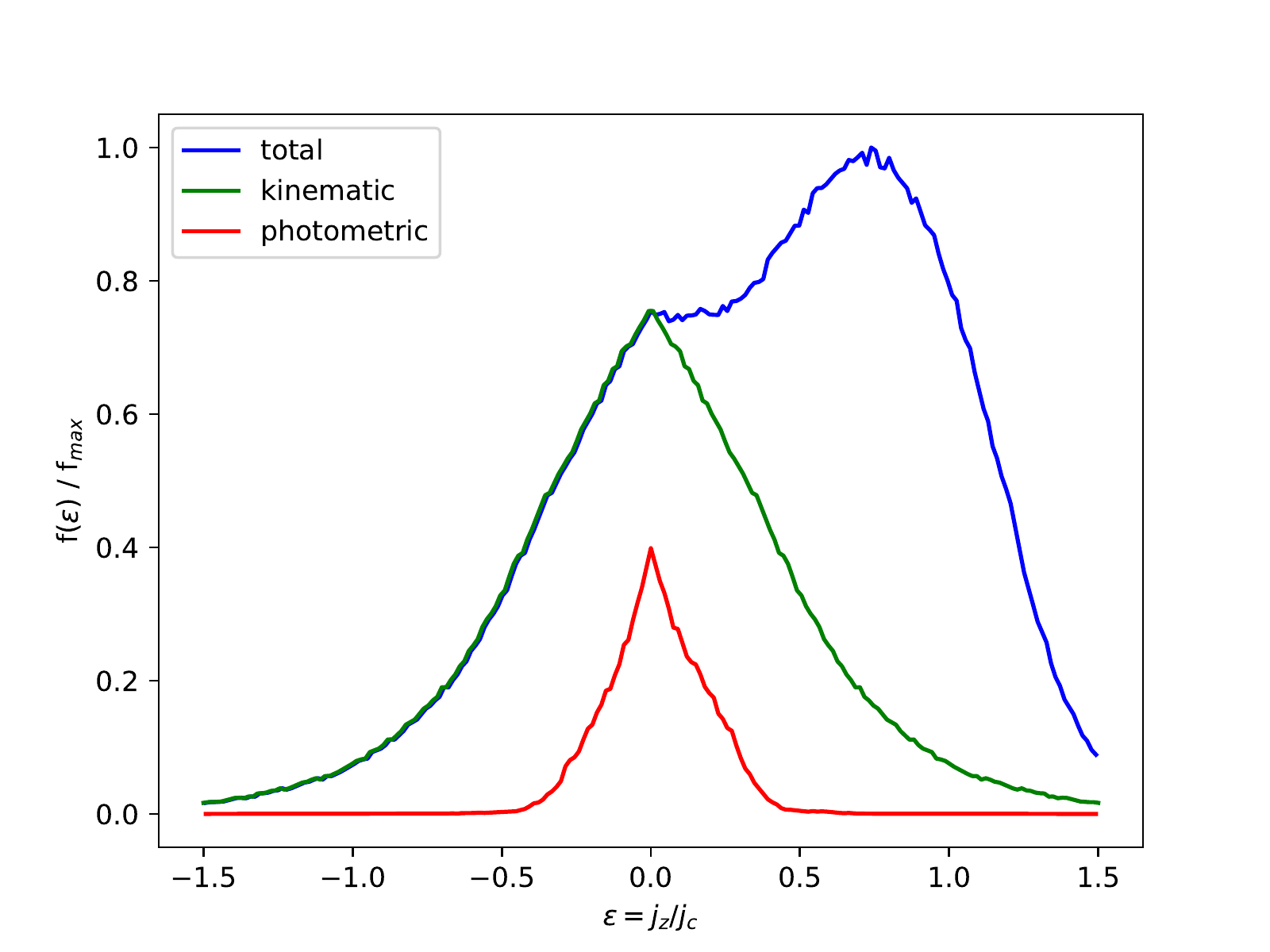}
\includegraphics[width=0.58\textwidth]{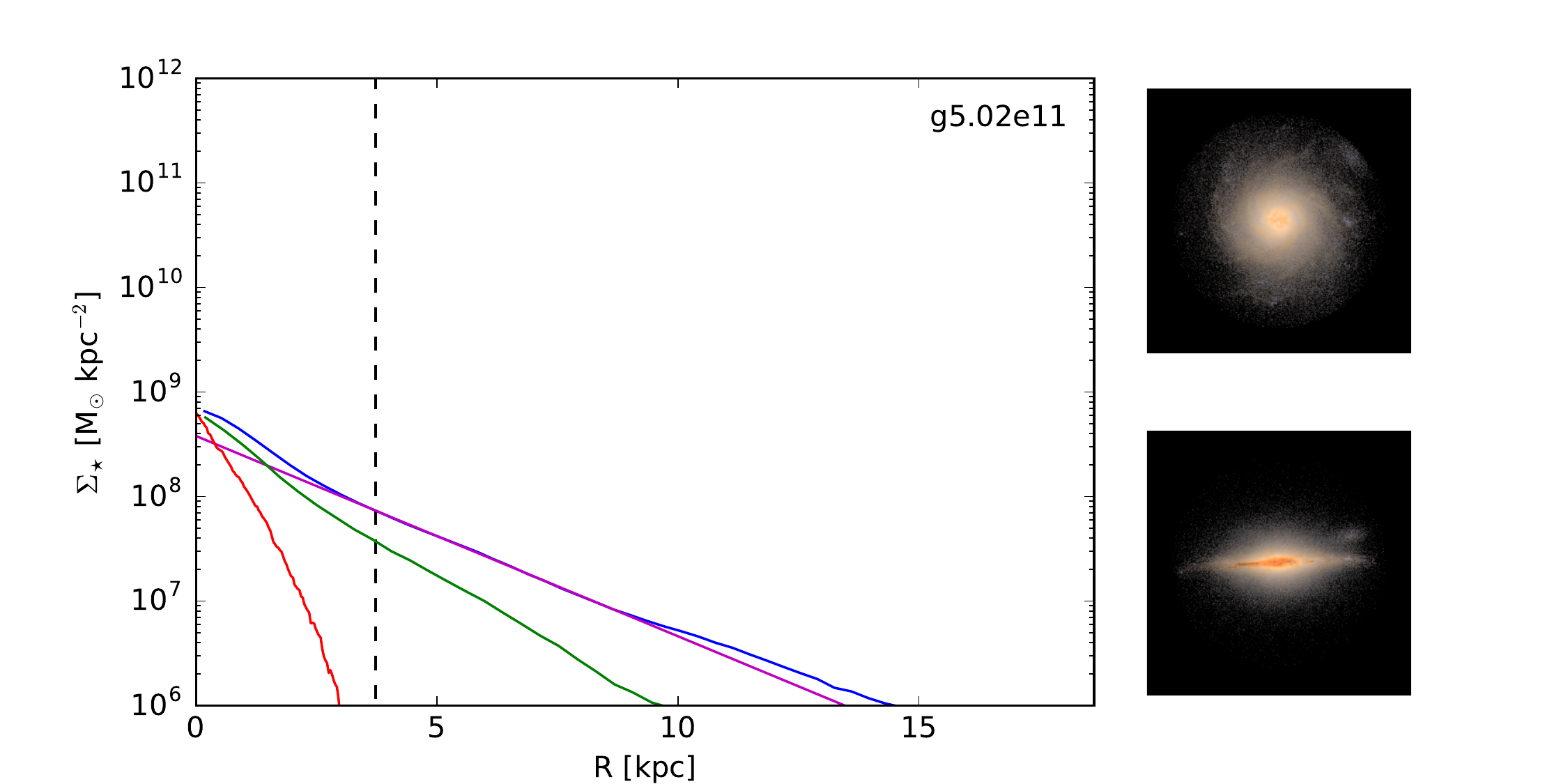}
\caption{Same as Fig.~\ref{fig:pro_img}, but for galaxies g3.49e11, g3.71e11 and g5.02e11.
         \label{fig:pro_img_append}}
\end{figure*}

\begin{figure*}
\includegraphics[width=0.4\textwidth]{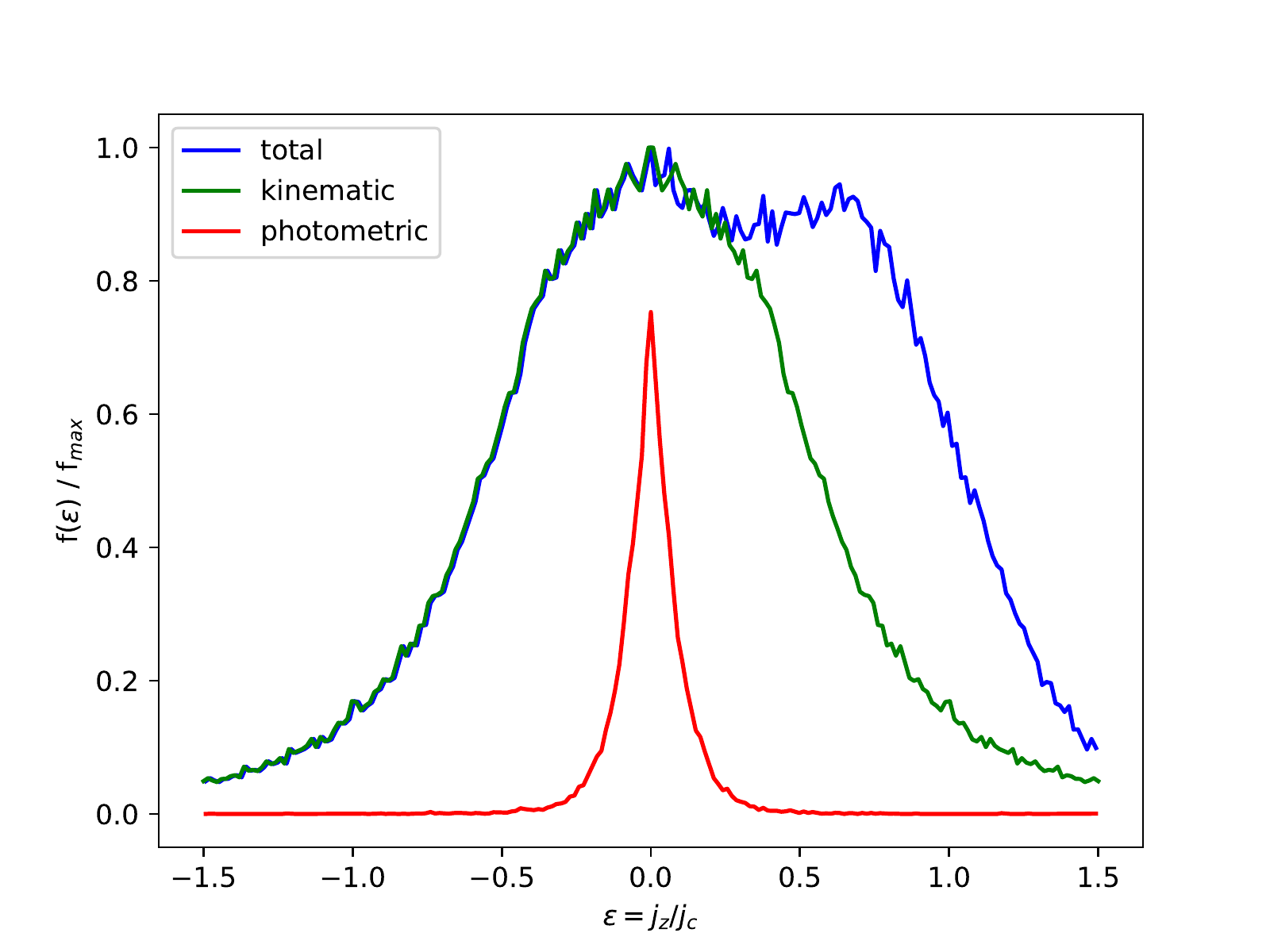}
\includegraphics[width=0.58\textwidth]{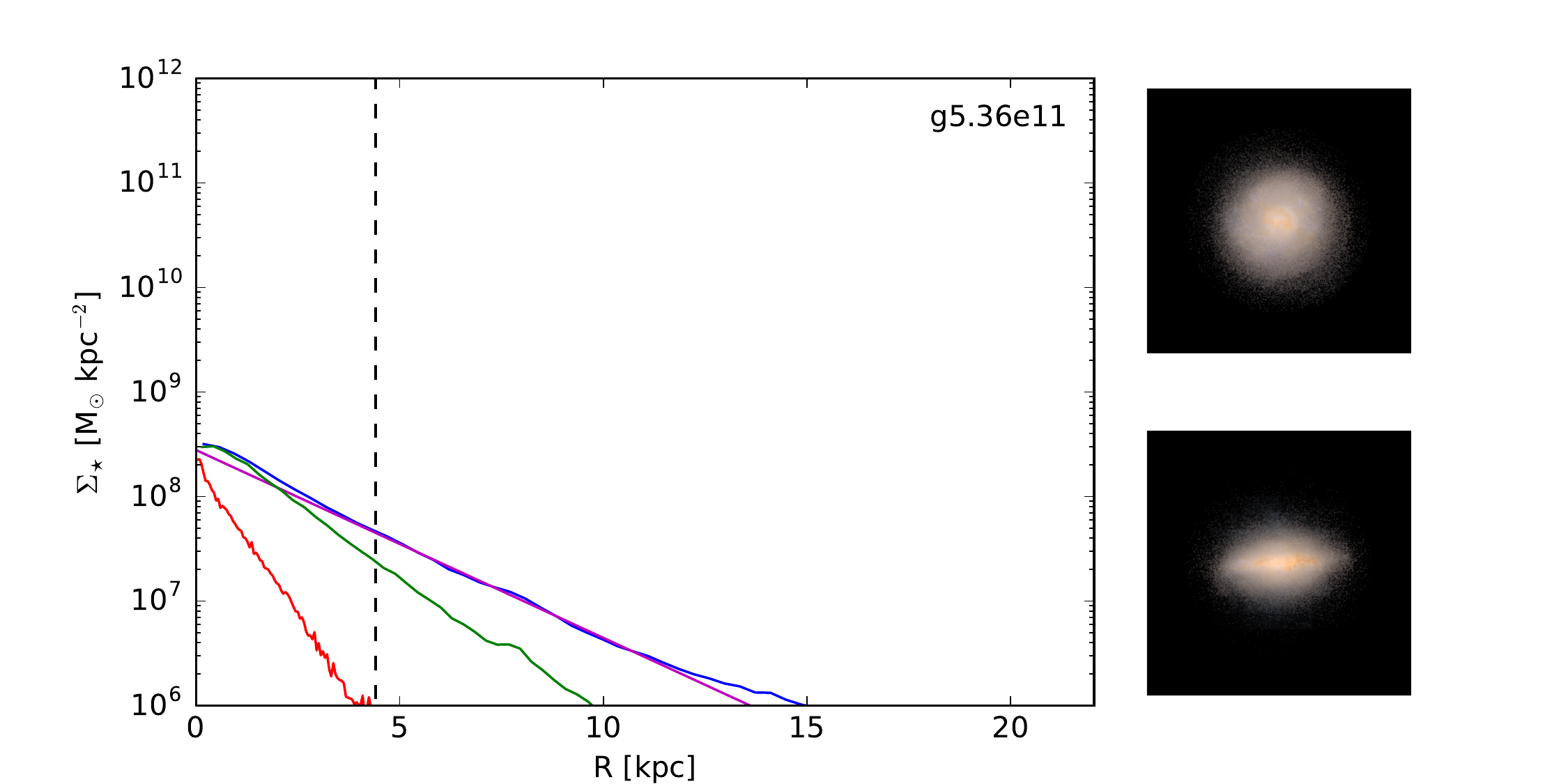}
\includegraphics[width=0.4\textwidth]{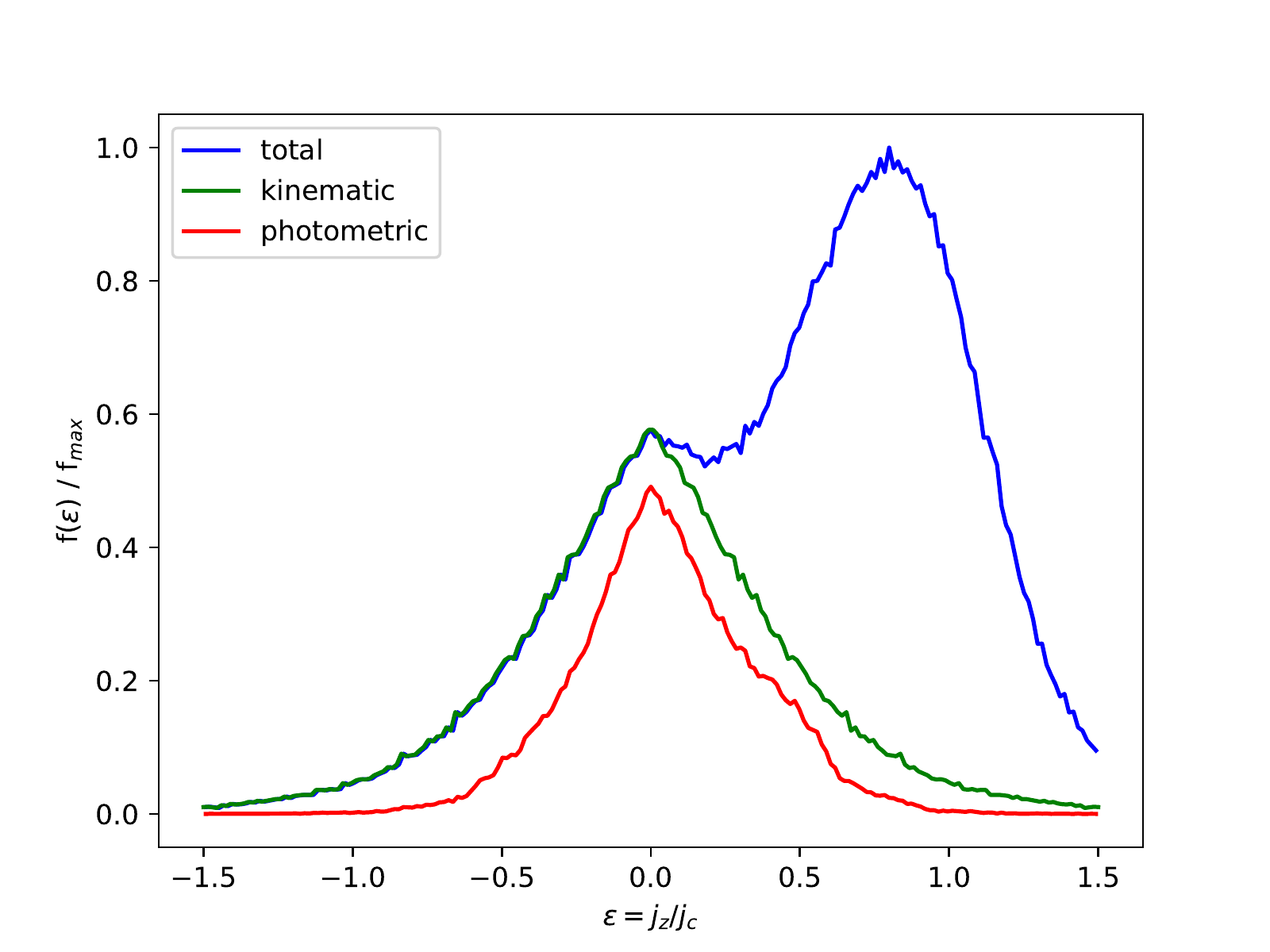}
\includegraphics[width=0.58\textwidth]{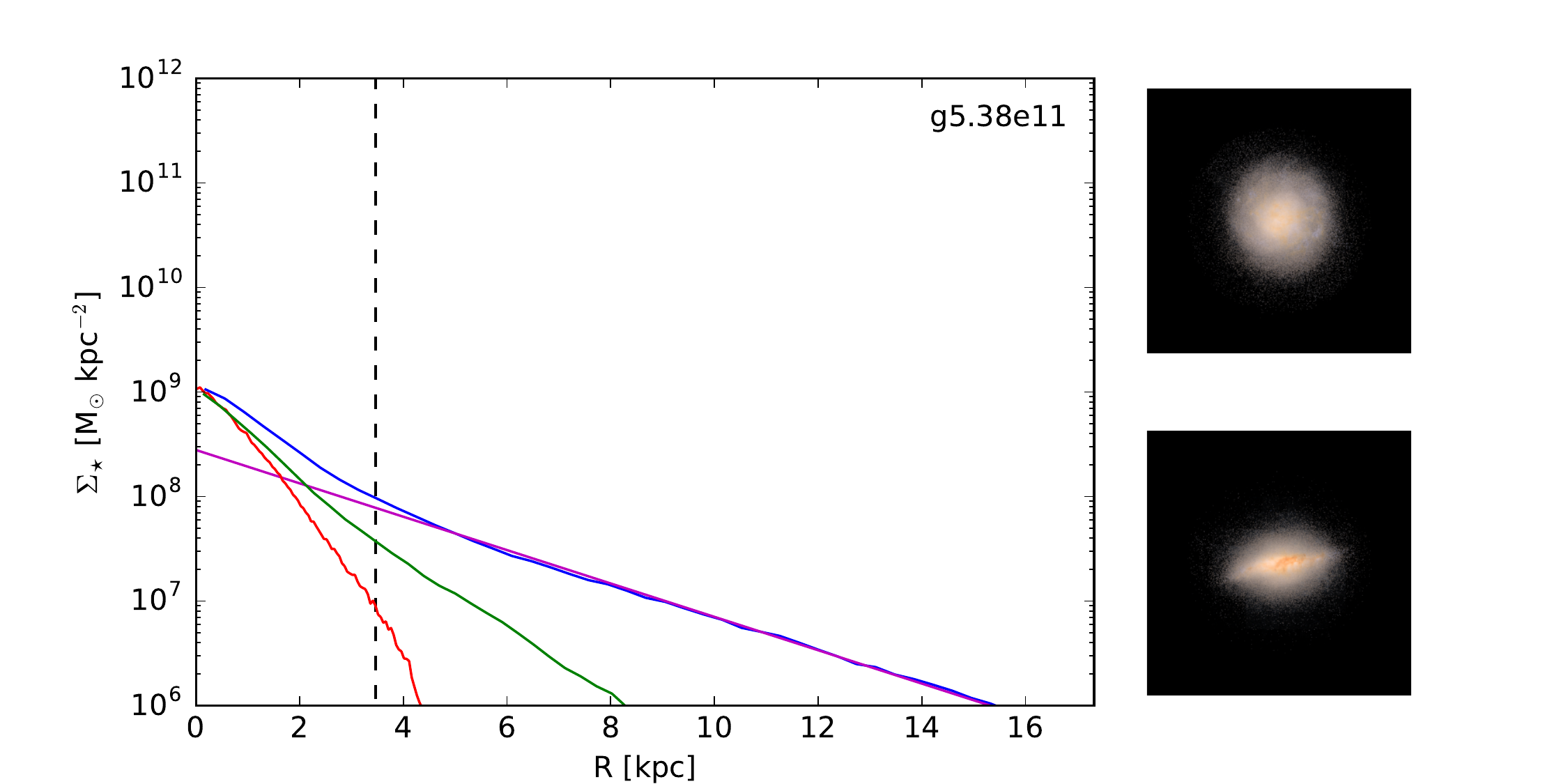}
\includegraphics[width=0.4\textwidth]{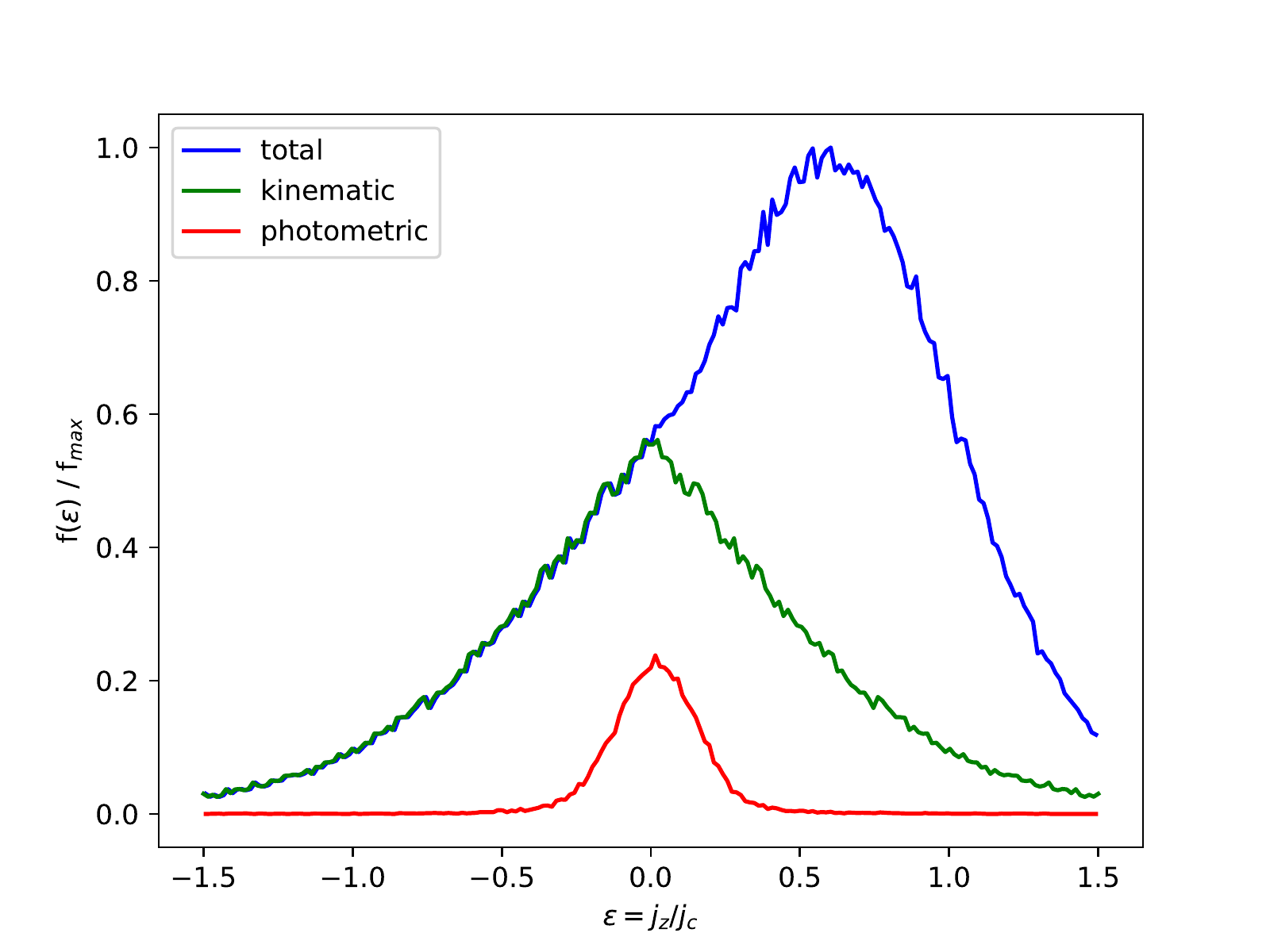}
\includegraphics[width=0.58\textwidth]{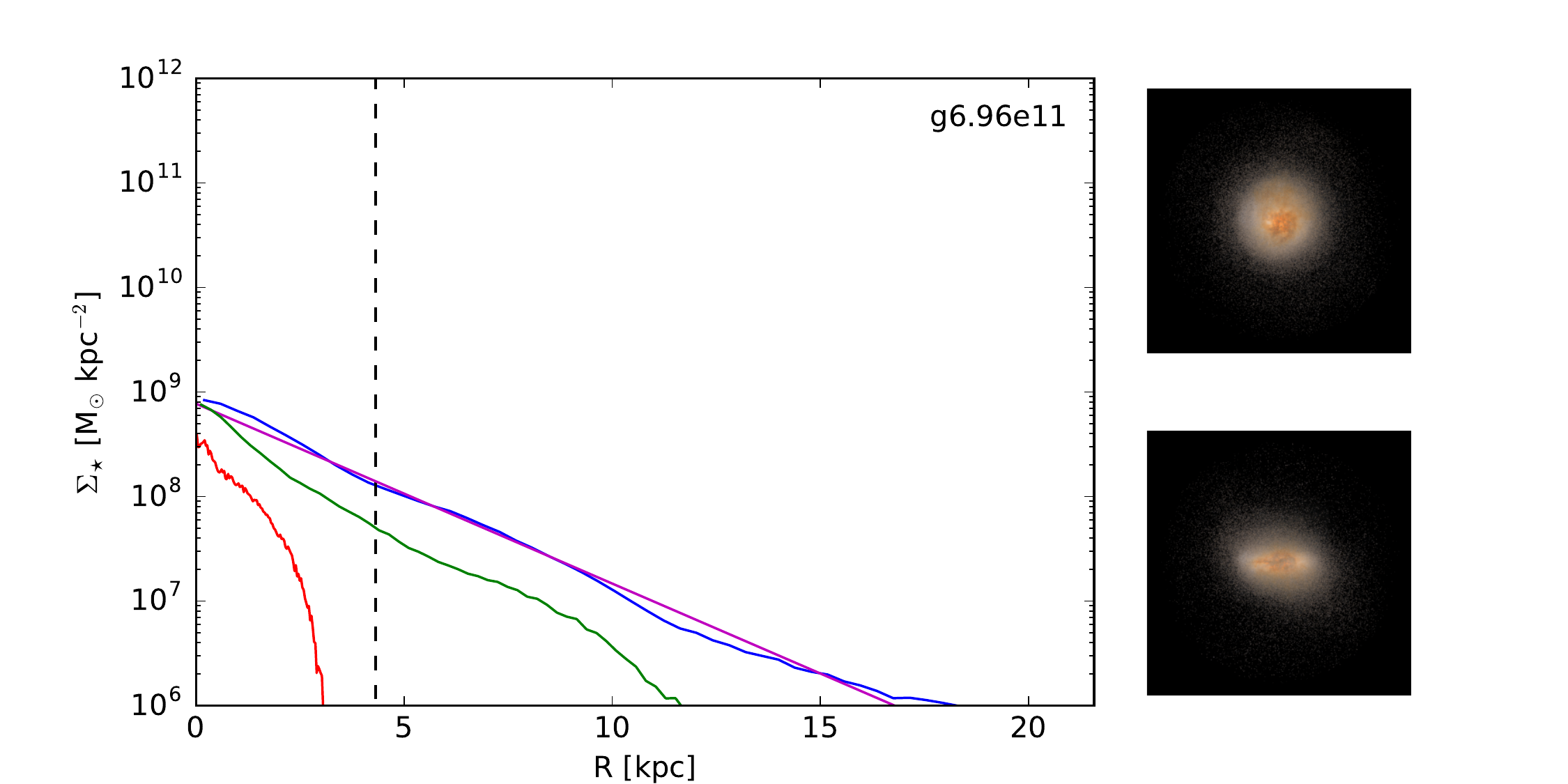}
\includegraphics[width=0.4\textwidth]{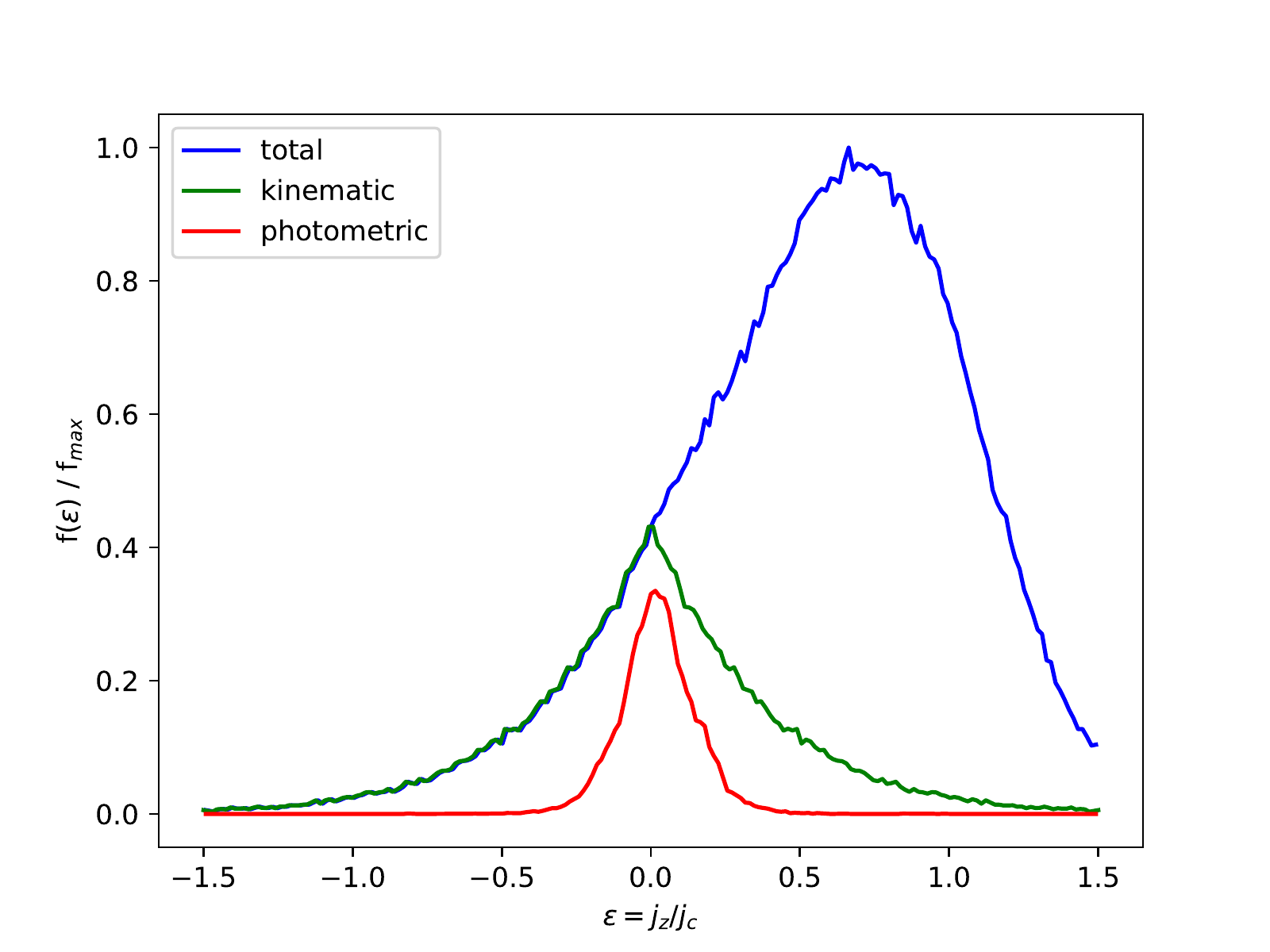}
\includegraphics[width=0.58\textwidth]{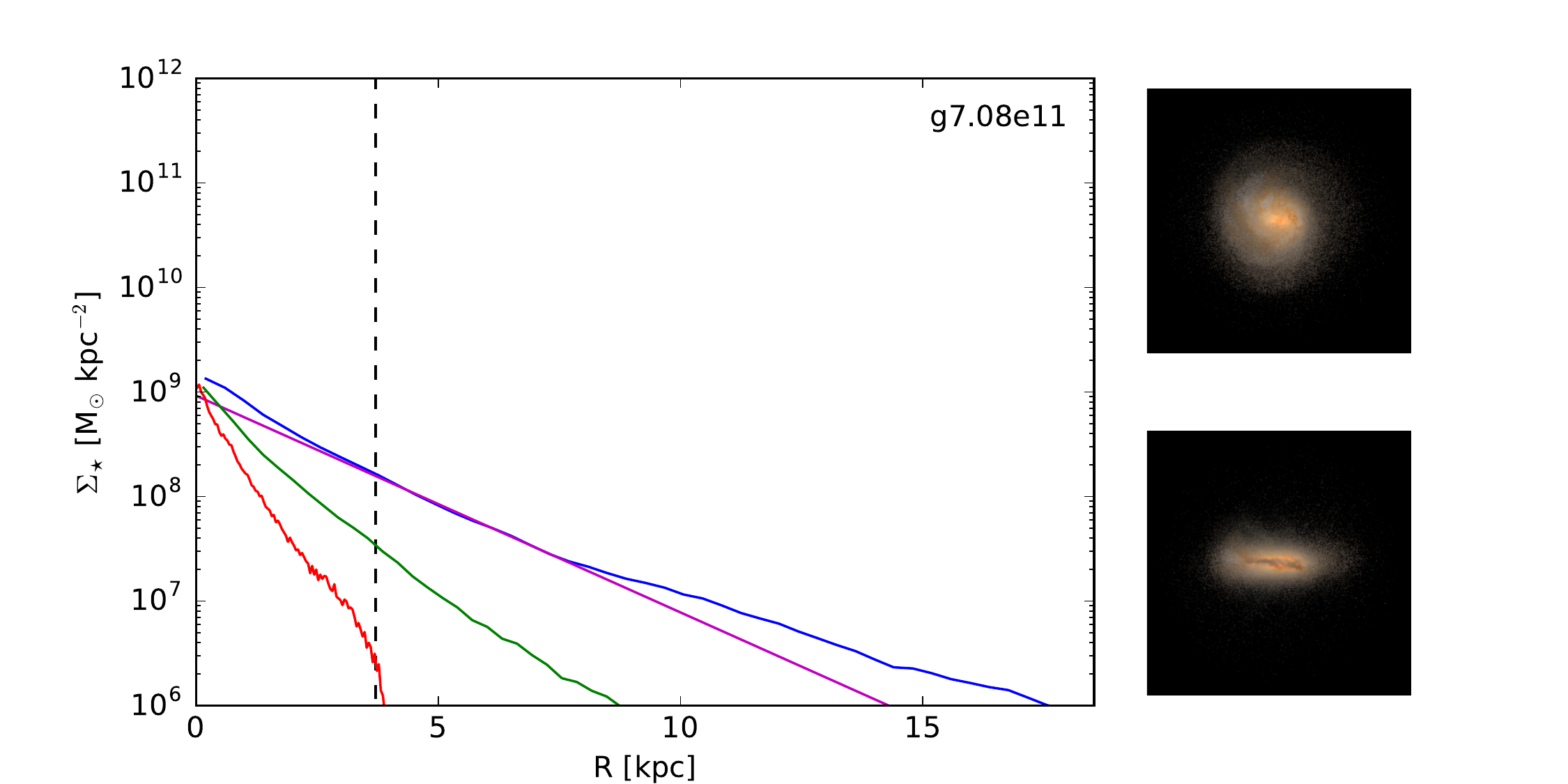}
\caption{Same as Fig.~\ref{fig:pro_img}, but for galaxies g5.36e11, g5.38e11, g6.96e11 and g7.08e11.
         \label{fig:pro_img_append}}
\end{figure*}

\begin{figure*}
\includegraphics[width=0.4\textwidth]{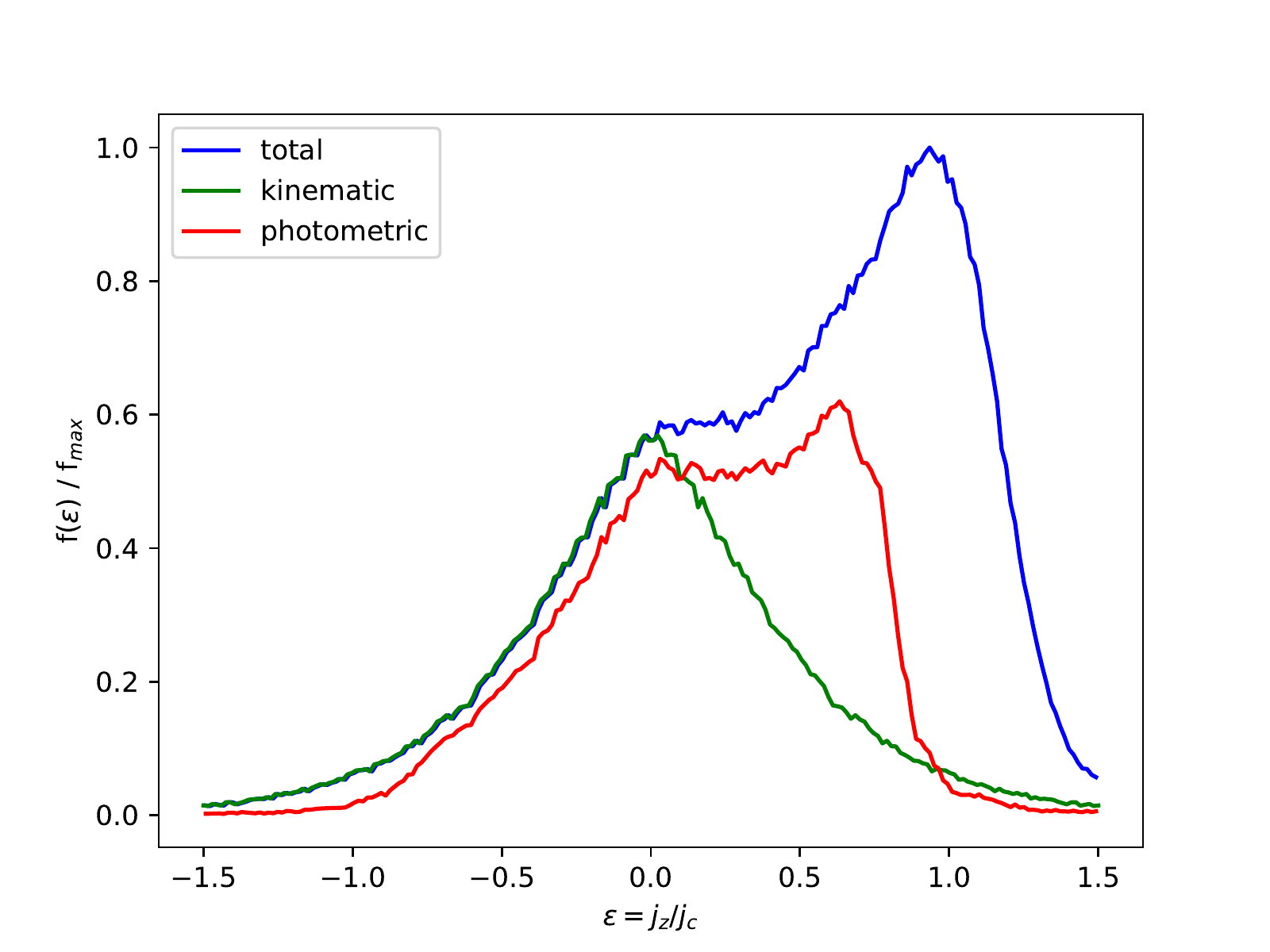}
\includegraphics[width=0.58\textwidth]{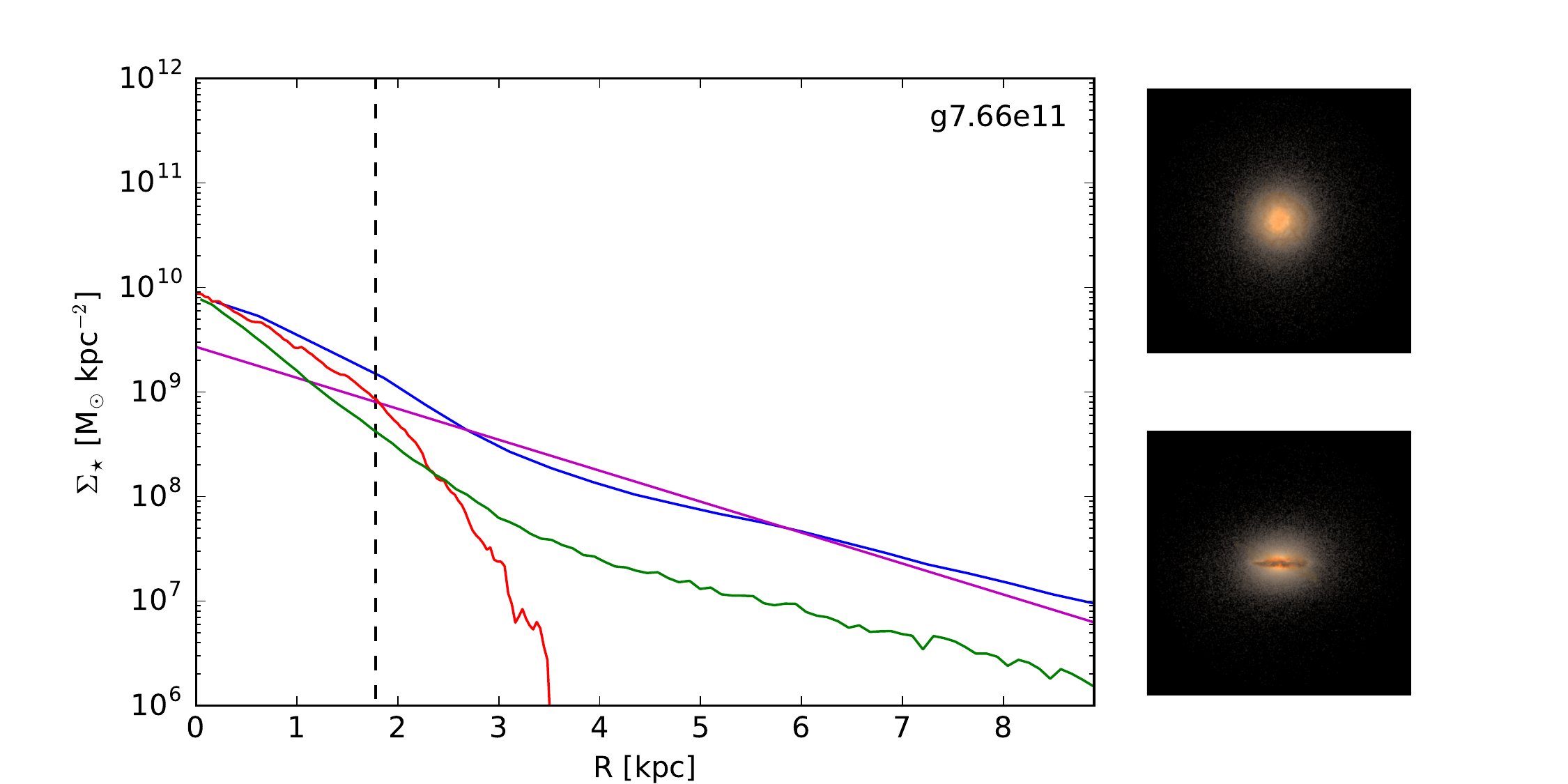}
\includegraphics[width=0.4\textwidth]{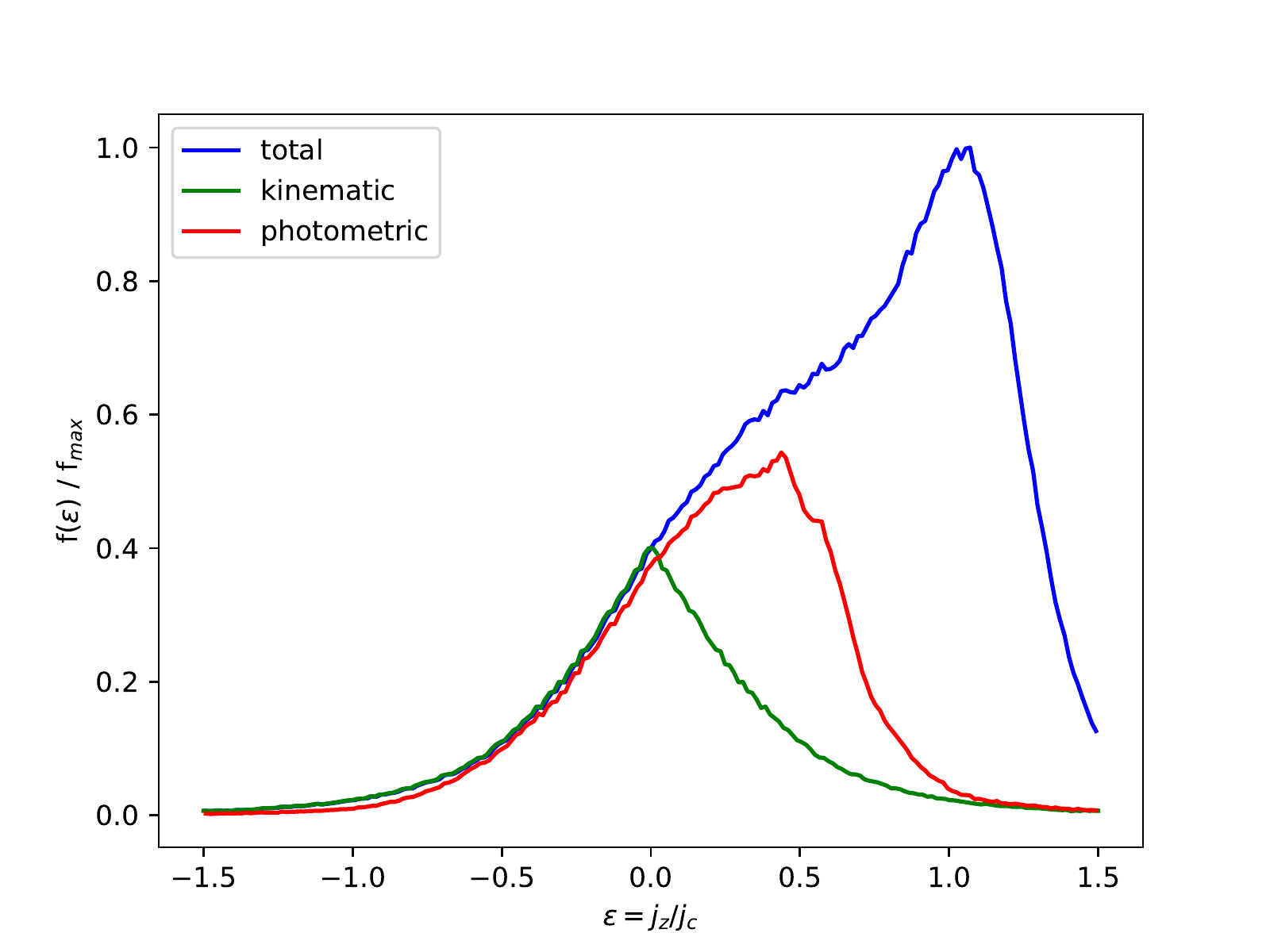}
\includegraphics[width=0.58\textwidth]{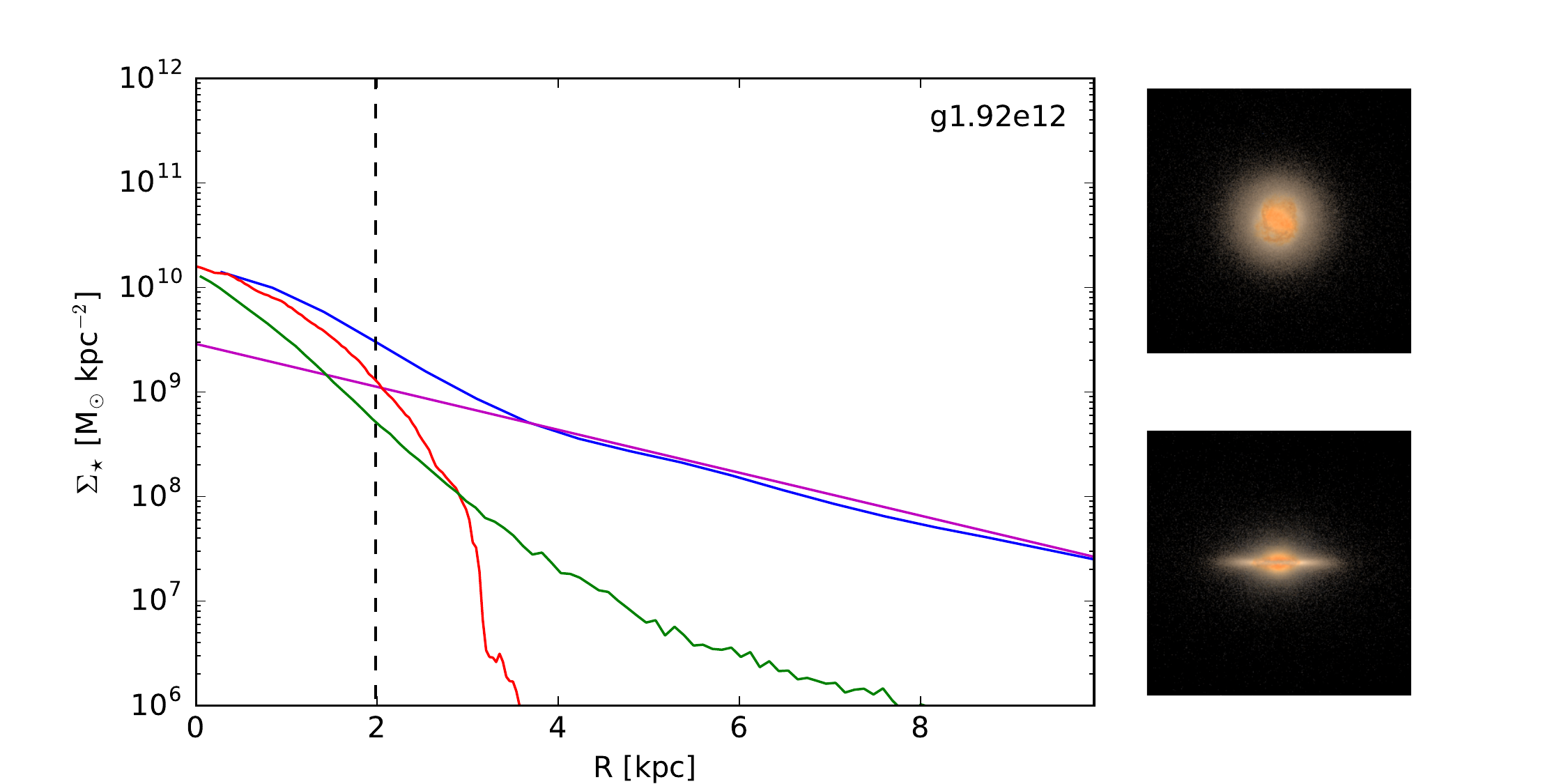}
\caption{Same as Fig.~\ref{fig:pro_img}, but for galaxies g7.66e11 and g1.92e12.
         \label{fig:pro_img_append}}
\end{figure*}

\bsp	
\label{lastpage}
\end{document}